\documentclass[conference]{IEEEtran}
\usepackage{cite}
\usepackage{amsmath,amssymb,amsfonts}
\usepackage{algorithmic}
\usepackage{graphicx}
\usepackage{textcomp}
\usepackage{xcolor}
\usepackage{fancyhdr}
\usepackage[hyphens]{url}

\def\BibTeX{{\rm B\kern-.05em{\sc i\kern-.025em b}\kern-.08em
    T\kern-.1667em\lower.7ex\hbox{E}\kern-.125emX}}

\usepackage{color,xcolor}
\usepackage{xspace}
\usepackage{multirow}
\usepackage{booktabs}
\usepackage{lipsum}
\usepackage{soulutf8}

\usepackage{colortbl}

\usepackage{amsmath}
\usepackage{amsfonts}
\usepackage{url}

\usepackage{afterpage}

\usepackage[ruled,vlined]{algorithm2e}
\usepackage{setspace}

\usepackage{pifont}

\usepackage{titlesec}

\usepackage{hyperref}
\hypersetup{
    colorlinks=true,
    linkcolor=magenta,
    filecolor=magenta,      
    urlcolor=blue,
}

\definecolor{citecolor}{RGB}{34,139,34}
\definecolor{mydarkblue}{rgb}{0,0.08,1}
\definecolor{mydarkgreen}{rgb}{0.02,0.6,0.02}
\definecolor{mydarkred}{rgb}{0.8,0.02,0.02}
\definecolor{mydarkorange}{rgb}{0.40,0.2,0.02}
\definecolor{mypurple}{RGB}{111,0,255}
\definecolor{myred}{rgb}{1.0,0.0,0.0}
\definecolor{mygold}{rgb}{0.75,0.6,0.12}
\definecolor{myblue}{rgb}{0,0.2,0.8}
\definecolor{mydarkgray}{rgb}{0.,0.2,0.2}

\definecolor{lightred}{RGB}{255,235,235}
\definecolor{lightgreen}{RGB}{235,255,235}
\definecolor{lightblue}{RGB}{235,235,255}
\definecolor{lightcyan}{RGB}{235,255,255}
\definecolor{lightmagenta}{RGB}{255,235,255}
\definecolor{lightyellow}{RGB}{255,255,235}

\definecolor{qxkcolor}{RGB}{215,235,255}
\definecolor{softmaxcolor}{RGB}{230,235,255}
\definecolor{probxvcolor}{RGB}{255,255,235}

\definecolor{topkcolor}{RGB}{255,235,235}
\definecolor{zecolor}{RGB}{255,255,235}
\definecolor{dynacolor}{RGB}{235,255,255}

\definecolor{reviewcolor}{RGB}{0,0,200}

\renewcommand\footnotemark{}

\newcommand{\etc}{{etc.}\xspace}
\newcommand{\ie}{i.e., }

\newcommand{\name}{SpAtten\xspace}
\newcommand{\spatten}{SpAtten\xspace}

\newcommand{\x}{$\times$\xspace}

\newcommand{\topk}{top-k\xspace}

\newcommand{\sota}{state-of-the-art\xspace}
\newcommand{\softmax}{Softmax\xspace}
\newcommand{\codesign}{co-design\xspace}

\newcommand{\flops}{FLOPs\xspace}
\newcommand{\bitwidth}{bitwidth\xspace}
\newcommand{\bitwidths}{bitwidths\xspace}
\newcommand{\lowprec}{quantization\xspace}

\newcommand{\athree}{{$A^3$}\xspace}
\newcommand{\mnnfast}{MNNFast\xspace}
\newcommand{\bert}{{BERT}\xspace}
\newcommand{\gpttwo}{{GPT-2}\xspace}

\newcommand{\titanxp}{TITAN\xspace Xp\xspace GPU}
\newcommand{\nano}{Nano\xspace GPU}
\newcommand{\xeon}{Xeon\xspace CPU}
\newcommand{\rasp}{Raspberry Pi ARM\xspace CPU}

\newcommand{\gflops}{GFLOPs\xspace}
\newcommand{\tflopss}{TFLOPS\xspace}
\newcommand{\gflopss}{GFLOPS\xspace}

\newcommand{\spattenfull}{SpAtten}

\newcommand{\spattensmall}{SpAtten$_{\mathrm{1/8}}$}
\newcommand{\spattenfc}{SpAtten-e2e}
\newcommand{\squad}{SQuAD\xspace}

\newcommand{\powerbert}{PoWER-BERT}

\newcommand{\numbenchmark}{30\xspace}

\newcommand{\prunedramsaving}{1.9}
\newcommand{\hpdramsaving}{1.1}
\newcommand{\dynadramsaving}{5.1}

\newcommand{\prunegpttwodramsaving}{3.8}

\newcommand{\spattenbertflops}{1.61}
\newcommand{\spattengpttwoflops}{0.43}

\newcommand{\gpubertflops}{0.02}
\newcommand{\gpugpttwoflops}{0.01}

\newcommand{\totaldramsaving}{10.0}
\newcommand{\totalcompsaving}{2.1}

\newcommand{\areaspattenfull}{18.71}

\newcommand{\topkarearatio}{2.7\%}

\newcommand{\powercompute}{1.36}
\newcommand{\powersram}{1.24}
\newcommand{\powerdram}{5.71}
\newcommand{\powerspattenfull}{8.30}

\newcommand{\topkpowerratio}{1.0\%}

\newcommand{\perfovertitanxp}{162} 
\newcommand{\perfoverxeon}{347}
\newcommand{\perfovernano}{1095}
\newcommand{\perfoverrasp}{5071}

\newcommand{\perfoverathree}{1.6}
\newcommand{\perfovermnnfast}{3.0}

\newcommand{\perffcovertitanxptwelve}{24} 
\newcommand{\perffcoverxeontwelve}{83}
\newcommand{\perffcovertitanxpteight}{35} 
\newcommand{\perffcoverxeoneight}{122}

\newcommand{\eeovertitanxp}{1193}
\newcommand{\eeoverxeon}{4059}
\newcommand{\eeovernano}{406}
\newcommand{\eeoverrasp}{1910}

\newcommand{\eeoverathree}{1.4}
\newcommand{\eeovermnnfast}{3.2}

\newcommand{\aeoverathree}{2.2}

\newcommand{\cmark}{\ding{51}}%
\newcommand{\xmark}{\ding{55}}%

\newcounter{rlabelno}

\title{\name: Efficient \underline{Sp}arse \underline{Atten}tion Architecture with Cascade Token and Head Pruning
}

\author{\IEEEauthorblockN{Hanrui Wang, Zhekai Zhang, Song Han}
\IEEEauthorblockA{
\textit{Massachusetts Institute of Technology}\\
\textit{Cambridge, MA, US} \\
\textit{\{hanrui, zhangzk, songhan\}@mit.edu}}
\texttt{\url{https://github.com/mit-han-lab/spatten}}
}

\begin{document}
\maketitle
\pagestyle{plain}
\thispagestyle{fancy}


\begin{abstract}

The attention mechanism is becoming increasingly popular in Natural Language Processing (NLP) applications, showing superior performance than convolutional and recurrent architectures. 
However, attention becomes the computation bottleneck because of its quadratic computational complexity to input length, complicated data movement and low arithmetic intensity. 
Moreover, existing NN accelerators mainly focus on optimizing convolutional or recurrent models, and cannot efficiently support attention. In this paper, we present \name, an efficient \emph{algorithm-architecture \codesign} that leverages token sparsity, head sparsity, and \lowprec opportunities to reduce the attention computation and memory access.
Inspired by the high redundancy of human languages, we propose the novel \emph{cascade token pruning} to prune away unimportant tokens in the sentence. We also propose \emph{cascade head pruning} to remove unessential heads.
Cascade pruning is fundamentally different from weight pruning since there is no trainable weight in the attention mechanism, and the pruned tokens and heads are selected on the fly. To efficiently support them on hardware, we design a novel \emph{\topk engine} to rank token and head importance scores with high throughput. Furthermore, we propose \emph{progressive \lowprec} that first fetches MSBs only and performs the computation; if the confidence is low, it fetches LSBs and recomputes the 
attention outputs, trading computation for memory reduction.

Extensive experiments on {\numbenchmark} benchmarks show that, on average, \name reduces DRAM access by \totaldramsaving\x with no accuracy loss, and achieves \perfoverathree\x, \perfovermnnfast\x, \perfovertitanxp\x, \perfoverxeon\x speedup, and \eeoverathree\x, \eeovermnnfast\x, \eeovertitanxp\x, \eeoverxeon\x energy savings over \athree accelerator, \mnnfast accelerator, \titanxp, \xeon, respectively.

\end{abstract}

\begin{IEEEkeywords}
Natural Language Processing; Attention; Domain-Specific Accelerator; Algorithm-Architecture Co-design; Pruning; Quantization

\end{IEEEkeywords}

\section{Introduction}

Natural Language Processing (NLP) has witnessed rapid progress in recent years driven by the attention mechanism~\cite{46201}. 
Attention models such as Transformer~\cite{46201}, \bert~\cite{devlin2018bert}, and \gpttwo~\cite{radford2019language} provide significant performance improvements over models based on Convolutional Neural Networks (CNN) and Recurrent Neural Networks (RNN). \bert~\cite{devlin2018bert} even outstrips human performance on the challenging sentence classification~\cite{wang-etal-2018-glue} tasks. 

Unfortunately, the high accuracy is at the expense of efficiency. Attention runs extremely slow on general-purpose platforms, due to its quadratic computational complexity to input length, complex data movement and low arithmetic intensity. 
For instance, to generate a sentence with only 30 tokens, a \gpttwo model takes a total of 370ms on a \titanxp \xspace to perform attention inference. That is two orders of magnitude slower than MobileNet-V2, which takes only 6ms to classify an image. 
On resources-limited \rasp, attentions cost 43s, making interactive dialog applications impossible. The inefficiency is even more severe for modern Large Language Models (LLMs) with long context length and large KV cache.
The efficiency barrier prevents attention models from being deployed on mobile devices. 
Many accelerators have been proposed to accelerate CNN and RNN, but they cannot be easily applied to attention due to the distinct operations. 

\begin{figure}[t]
    \centering
    \includegraphics[width=\columnwidth]{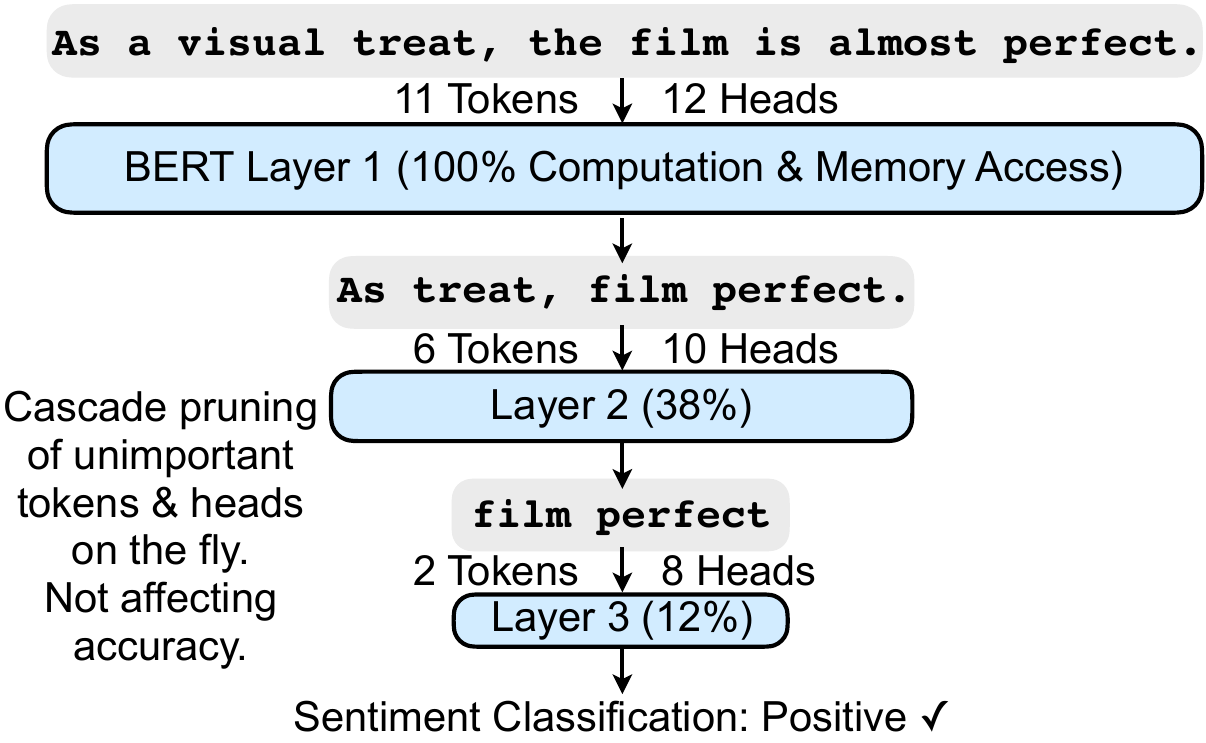}
    \vspace{-20pt}
    \caption{\textit{Cascade token and head pruning} removes redundant tokens and heads globally across layers. Evaluated with BERT-Base on SST-2 dataset.}
    \vspace{-10pt}
    \label{fig:teaser}
\end{figure}

In this paper, we propose \name\footnote{\name is homophonic with \emph{spartan}, meaning simple and frugal. It is analogous to \emph{token and head pruning}, making sentences shorter and simpler.}, an algorithm-architecture \codesign to enable efficient attention inference. We propose three algorithmic optimizations: \emph{cascade token pruning}, \emph{cascade head pruning} and \emph{progressive quantization} to reduce computation and memory access. Different from conventional techniques, pruning is applied to the tokens and heads, not weights. Cascade means: once a token/head is pruned, it is removed in all following layers, so one layer only needs to process remaining tokens/heads from previous layers. The deeper the layer, the more tokens/heads are pruned. The three techniques are input-dependent since the pruned computation and bit-width are \emph{adaptive} to input instances. Cascade pruning requires sorting token/head importance scores on the fly. Thus we design hardware architecture with high parallelism \topk engines for token/head selections, specialized memory hierarchy, and fully-pipelined datapath to translate theoretical savings to real speedup and energy reduction.

The inputs of attention contain Query (Q), Key (K), and Value (V), each split into multiple heads.
Then attention probabilities are computed as the softmax of Q $\times$ K. Multiplying the attention probabilities with V gives the result of one head; concatenating all heads together gives the attention output.
The arithmetic intensity of attention in the generation stage is low: only two operations per data (0.5ops/Byte) for the vector-matrix multiplication (Q $\times$ K). Generation takes the largest part of overall latency in \gpttwo models (97\% when generating 32 tokens); thus, the overall performance is memory-bounded. For \bert, the overall performance is computation-bounded.

Therefore, we propose \emph{cascade token pruning} as shown in Figure~\ref{fig:teaser} to reduce both DRAM access and computation. Inspired by human languages being highly redundant due to many structural and meaningless tokens such as prepositions, articles, and adverbs, we can safely remove unimportant tokens with little impact on the results. Moreover, attention uses many heads to capture various dependencies, but some of them are also redundant~\cite{voita2019analyzing}. Therefore, we also propose \emph{cascade head pruning} to determine the importance of heads based on their influence on the outputs, then prune unessential heads. Cascade token and head pruning are fundamentally different from the classic weight pruning and classic head pruning because: (i) There is no trainable weight in attention. (ii) Conventionally, pruned weights and heads are determined at compile-time and consistent for all inputs. In contrast, tokens and heads to be pruned in \name are selected on the fly and vary between different inputs. Token pruning is also different from classic activation pruning because it depends on attention probabilities, not activation magnitude. Specifically, we prune the tokens according to cumulative token importance scores obtained by accumulating attention probabilities (indicators for token influence) across layers. Since long sentences are naturally more redundant, we also adjust the pruning ratios based on sentence length: the longer, the more tokens are pruned away. Moreover, the heads are pruned according to cumulative head importance scores, which are computed by accumulating each head's magnitude across layers. 
To support cascade token and head pruning, we design and implement a specialized high parallelism \topk engine with $O(n)$ time complexity to get the $k$ most essential tokens or heads. On average, token and head pruning can reduce DRAM access and computation by \prunegpttwodramsaving\x and \hpdramsaving\x on eight \gpttwo models.

To further reduce the DRAM access, we also propose \emph{progressive \lowprec} for attention inputs. We find an interesting phenomenon that quantization errors are related to attention probability distributions: if a few tokens dominate the distribution, the quantization error is small -- only MSB is needed; for a flat distribution, the error is large -- both LSB and MSB are needed. 
We also provide a theoretical proof for this phenomenon in Section~\ref{sec:method:dyna}. 
Based on this observation, we quantize more aggressively for attention with dominated attention probabilities and more conservatively for others. Concretely, we first fetch MSBs of attention inputs to compute the attention probabilities. If the max probability is smaller than a threshold, indicating the distribution is flat, we will fetch LSBs on-chip and recompute attention probabilities. 
In such a way, we trade computation to less memory access, which is beneficial to memory-bounded models. With progressive \lowprec, we can save another \dynadramsaving\x memory access.

Previous \sota attention accelerators $A^3$~\cite{ham20203} and \mnnfast~\cite{8980322} also leverage sparsity. However, they have three main limitations. 
(i) $A^3$ and \mnnfast need to fetch everything from DRAM before calculating what can be pruned. Thus the overhead is already paid, and no DRAM access is reduced. They only optimize computation-bounded discriminate models, and cannot accelerate memory-bounded generative models. \name not only improves computation-bounded discriminative ones (\bert), but also solves the challenge of memory-bounded generative ones (\gpttwo). It significantly reduces QKV DRAM access with token pruning (\prunegpttwodramsaving$\times$), head pruning (\hpdramsaving$\times$), and progressive quantization (\dynadramsaving$\times$). 
(ii) Head sparsity is an opportunity to further reduce DRAM access and computation, but \athree and \mnnfast do not support head pruning.
(iii) \athree prunes QKV vectors of a token only locally in one head, and \mnnfast only prunes V vector locally. Therefore they only reduce the attention layer's computation but not Feed-Forward Network (FFN) layers. \name prunes the token \emph{globally}: once a token is pruned, the involved computations in \emph{all} following layers are skipped. Therefore, the computations in FFN layers are also reduced in \name.

\name has a wide support range thanks to the generalization ability of attention-based NLP models. For instance, \bert can be used for arbitrary discriminative tasks, such as sentence sentiment classification and sentence similarity regression, for which the backbone of \bert is the same and only the last layer needs to be changed. Likewise, \gpttwo can handle all generative tasks, including language modeling, document summarization, \etc

\begin{figure}[t]
    \centering
    \includegraphics[width=1\columnwidth]{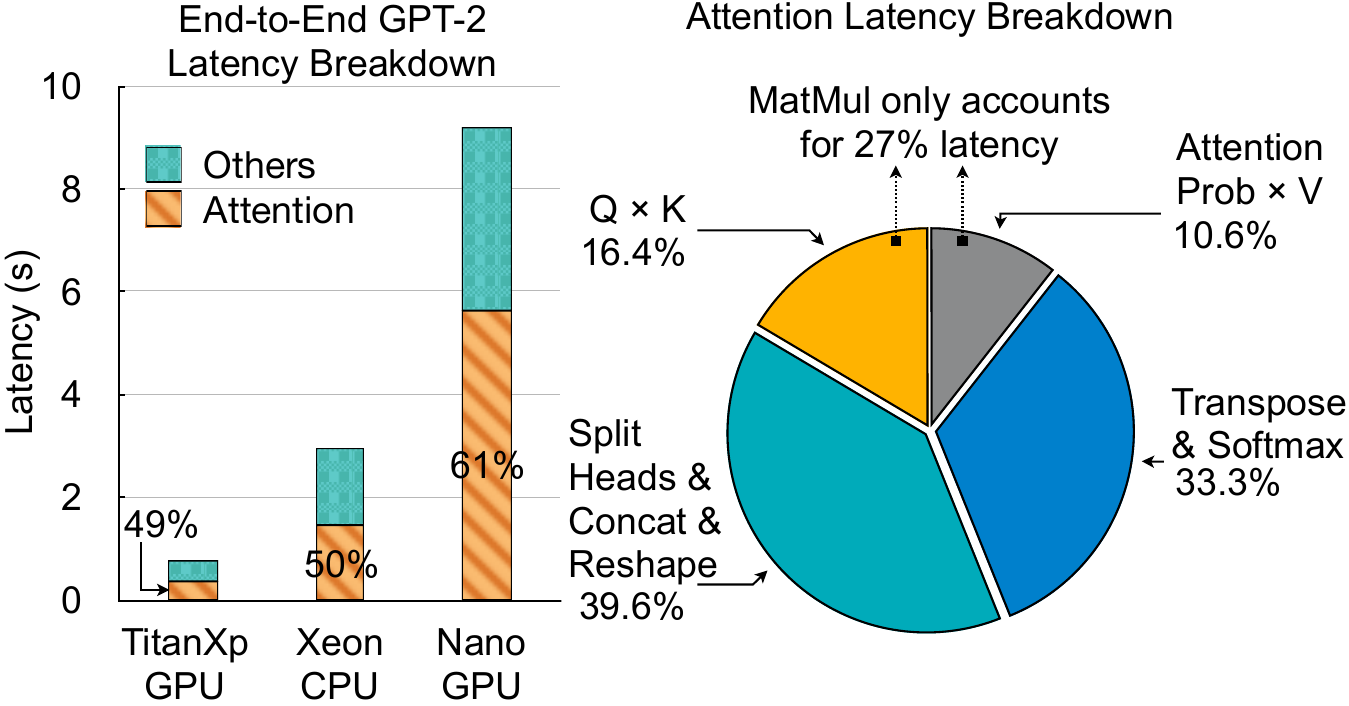}
    \vspace{-10pt}
    \caption{End-to-End \gpttwo latency breakdown on various platforms, and attention latency breakdown on \titanxp. Attention accounts for over 50\% of total latency. Data movements account for 73\% of attention latency.}
    \vspace{-15pt}
    \label{fig:latency_breakdown}
\end{figure}

In summary, \name performs algorithm-architecture \codesign for sparse and quantized attention computing while preserving the accuracy. It makes four contributions: 
\begin{itemize}
    \item \textbf{Cascade Token Pruning} removes unimportant tokens according to the cumulative token importance scores, reducing DRAM access and computation by up to \prunegpttwodramsaving\x. 
    \item \textbf{Cascade Head Pruning} removes unimportant heads and save DRAM access and computation by another \hpdramsaving\x.
    \item \textbf{Progressive Quantization} trades a little more computation for less memory access. We change the \bitwidths of different attention heads and layers based on attention probability distribution, reducing DRAM access by \dynadramsaving\x.
    \item \textbf{Specialized High Parallelism top-k Engine} with $O(n)$ time complexity to efficiently support on-the-fly token and head selections.
\end{itemize}

We extensively evaluate \name on \numbenchmark benchmarks including GLUE set~\cite{wang-etal-2018-glue}, SQuAD\cite{rajpurkar2016squad}, Wikitext-2~\cite{merity2016pointer}, Wikitext-103~\cite{merity2016pointer}, Pen Tree Bank~\cite{marcus-etal-1993-building} and Google One-Billion Word~\cite{chelba2013one} with \bert and \gpttwo models. \name reduces DRAM access by \totaldramsaving\x with \emph{no accuracy loss}, and achieves \perfoverathree\x, \perfovermnnfast\x, \perfovertitanxp\x, \perfoverxeon\x, \perfovernano\x, \perfoverrasp\x speedup, and \eeoverathree\x, \eeovermnnfast\x, \eeovertitanxp\x, \eeoverxeon\x, \eeovernano\x, \eeoverrasp\x energy savings over \athree accelerator, \mnnfast accelerator, \titanxp, \xeon, \nano, \rasp, respectively.

\section{Background and Motivation}

\begin{figure}[t]
    \centering
    \includegraphics[width=\columnwidth]{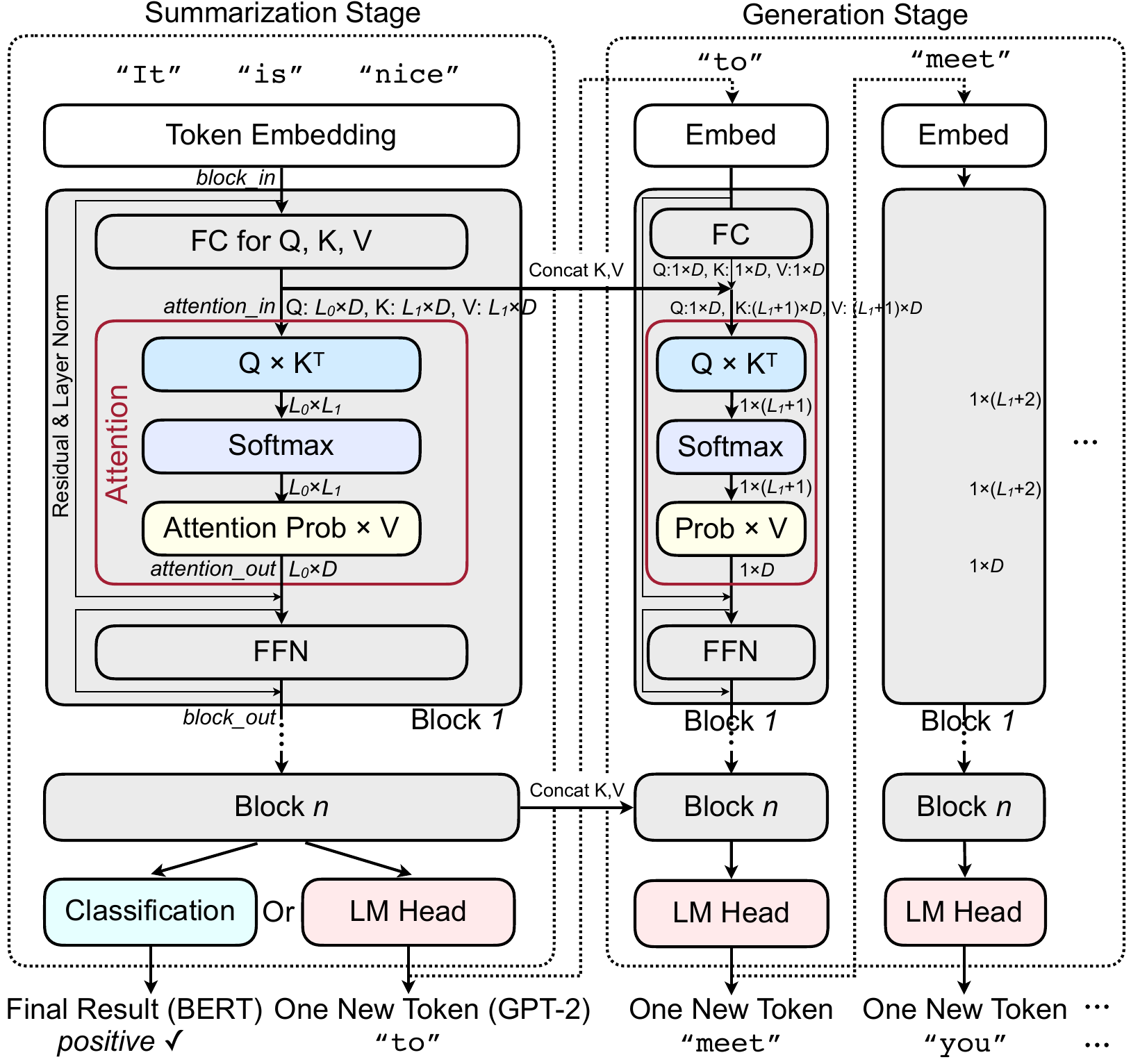}
    \vspace{-15pt}
    \caption{NLP model architecture with attention. \bert only contains the summarization stage. \gpttwo contains summarization and generation stages.
}
    
    \label{fig:background}
\end{figure}

\newlength{\oldtextfloatsep}\setlength{\oldtextfloatsep}{\textfloatsep}

\setlength{\textfloatsep}{0pt}
\begin{algorithm}[!t]
\SetKwInOut{Input}{Input}
\footnotesize{
    \SetAlgoLined
    \textbf{Input:} 
    $Q_{in} \in \mathbb{R}^{L_0 \times D_{in}}$,
    $K_{in} \in \mathbb{R}^{L_1 \times D_{in}}$, 
    $V_{in} \in \mathbb{R}^{L_1 \times D_{in}}$; \\ 
    Number of Heads: $h$ ; \\
    Split $Q_{in}, K_{in}, V_{in}$ to $h$ chunks:  \\
    $
    Q \in \mathbb{R}^{ h \times L_0 \times D}, K \in \mathbb{R}^{ h \times L_1 \times D},  V \in \mathbb{R}^{ h \times L_1 \times D}, D = \frac{D_{in}}{h} $; \\
    \For{$head_{id} = 0 \mathbf{ \ to \ } h$}{
    
    \colorbox{qxkcolor}{\vbox{ 
    $attention\_score \in \mathbb{R}^{  L_0 \times L_1}$ ;\\ 
    $attention\_score = Q[head_{id}] \cdot K[head_{id}]^T $ ;\\
    $attention\_score = attention\_score /$sqrt($D$) ; \\
    }}
    
    \colorbox{softmaxcolor}{\vbox{ 
    $attention\_prob \in \mathbb{R}^{  L_0 \times L_1}$ ;\\
    \For{$row_{id} = 0 \mathbf{ \ to \ } L_0$}{
    $attention\_prob[row_{id}] = $ Softmax($attention\_score[row_{id}]$)
    }
    }}
    
    \colorbox{probxvcolor}{\vbox{ 
    $E[head_{id}] = attention\_prob \cdot V[head_{id}]$ ;\\
    }}
    
    }
    
    Concatenate heads of $E \in \mathbb{R}^{ h \times L_0 \times D}$ as output; \\
    \textbf{Output:} $attention\_out \in \mathbb{R}^{L_0 \times D_{in}}$;\\
    
    \caption{Attention}
    }
    \label{algo:attention}

\afterpage{\global\setlength{\textfloatsep}{\oldtextfloatsep}}

\end{algorithm}

\subsection{Background}
\textbf{Attention-Based NLP Models.}
NLP tasks can be categorized into two types: discriminative and generative. For discriminative ones, the models need to summarize the input information and make predictions. Discriminative tasks include token-level classification\cite{tjong-kim-sang-de-meulder-2003-introduction}, sentence-level classification and regression~\cite{socher-etal-2013-recursive} \etc 
Meanwhile, models for generative tasks need first to summarize the input information and then generate new tokens. Exemplary generation tasks include Language Modeling (LM)~\cite{radford2019language} and machine translation~\cite{46201}.

\bert for discriminative and \gpttwo for generative tasks are the most widely-used models as illustrated in  Figure~\ref{fig:background}. \bert only contains the summarization stage, while \gpttwo contains summarization and generation stages. In summarization (Figure~\ref{fig:background} left), the input tokens are first embedded into vectors and processed by blocks. Inside each block, $block\_in$ are first multiplied with three matrices to get Query (Q), Key (K) and, Value (V). Then QKV are processed by attention to get the intermediate features $attention\_out$. A residual layer adds the $attention\_out$ with $block\_in$ and conducts layer normalization. There will be an additional FC on $attention\_out$ if there is more than one head. Furthermore, a Feed-Forward Network (FFN) layer containing two Fully-Connected (FC) layers is applied. Finally, another residual operation is conducted and outputs $block\_out$.
The same block is repeated multiple times such as 12 times for \bert-Base. The last block is followed by one classification layer in \bert to get the \emph{final result}. In contrast, \gpttwo applies an LM head to generate one new token and enters the generation stage.

Generation stage (Figure~\ref{fig:background} right) has two main differences from summarization: 1) Each iteration only processes \emph {one single token} instead of the whole sentence. 2) Ks and Vs from the summarization stage are concatenated with current K and V, and sent to attention \emph{in batch}, while the query is still \emph{one single vector}. After the last block, another new token will be generated. Generation stage ends when \texttt{end\_of\_sentence} token is generated, or the sentence length reaches a pre-defined limit. One generation iteration's runtime is similar to the whole summarization stage on GPU, because the summarization is processed in batch.

\begin{figure*}[t]
    \centering
    \includegraphics[width=1\textwidth]{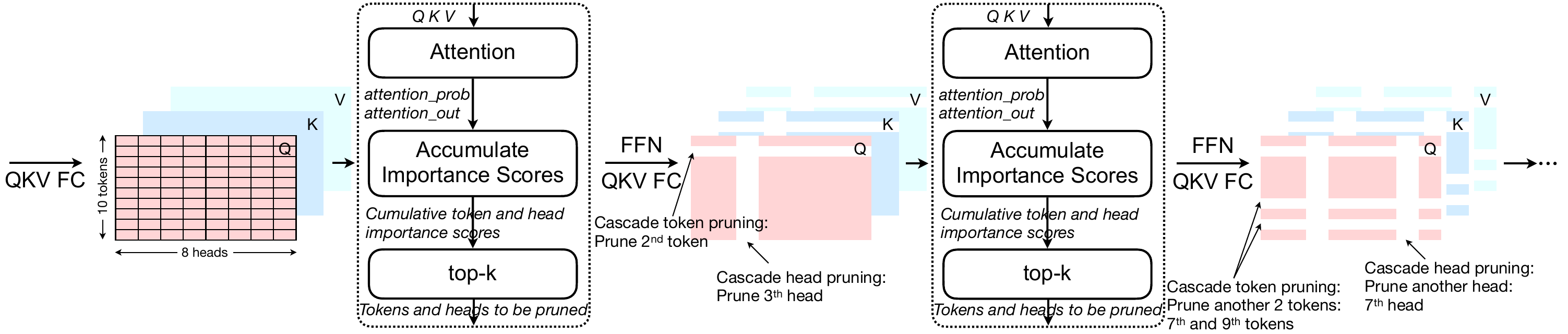}
    \vspace{-15pt}
    \caption{Cascade token pruning removes redundant tokens and corresponding entire Q K V vectors according to the cumulative token importance scores computed from $attention\_prob$. Cascade head pruning removes unimportant heads and corresponding chunks in all Q K V vectors according to the cumulative head important scores computed from $attention\_out$. Once a token/head is pruned, it will never appear in any following layers, thus named cascade pruning. More tokens and heads are pruned away as the layer goes deeper.}
    \vspace{-15pt}
    \label{fig:methodology}
\end{figure*}

\setlength{\textfloatsep}{0pt}

\begin{algorithm}[!t]
\footnotesize{
\SetKwInOut{Input}{Input}
    \SetAlgoLined
    \textbf{Input:} 
    Number of heads: $h$;\\
    Token pruning ratio: $p_{t}$; Head pruning ratio: $p_{h}$;\\
    $attention\_prob \in \mathbb{R}^{ h \times L_0 \times L_1}$, 
    $attention\_out \in \mathbb{R}^{L_0 \times D_{in}}$; \\
    Previous cumulative token importance score: $s_{t} \in L_1;$ \\
    Previous cumulative head importance score: $s_{h} \in h;$
    
    Reshape $attention\_out$ by head, get $E \in \mathbb{R}^{ h \times L_0 \times D}$;
    \tcc{accumulate the token importance}
    \For{$token_{id} = 0 \mathbf{ \ to \ } L_1$}{
        \For{$l_{0} = 0 \mathbf{ \ to \ } L_0$}{
            \For{$head_{id} = 0 \mathbf{ \ to \ } h$}{
                $s_{t}[token_{id}] \mathrel{{+}{=}} attention\_prob[head_{id}][l_0][token_{id}]$
            }
        }
    }
    \tcc{accumulate the head importance}
    \For{$head_{id} = 0 \mathbf{ \ to \ } h$}{
        \For{$l_{0} = 0 \mathbf{ \ to \ } L_0$}{
            \For{$d = 0 \mathbf{ \ to \ } D$}{
                $s_{h}[head_{id}] \mathrel{{+}{=}} abs(E[head_{id}][l_0][d] )$
            }
        }
    }
    
    $remained\_token\_id = $ top-k($s_t, L_1 \times (1-p_t)$) \\
    $remained\_head\_id = $ top-k($s_h, h \times (1-p_h)$) \\
    
    \textbf{Output:} $remained\_token\_id, remained\_head\_id, s_t, s_h$
    
    \caption{Token and Head Pruning (one layer)}
}
    \label{algo:cascade}
\afterpage{\global\setlength{\textfloatsep}{\oldtextfloatsep}}
\end{algorithm}

\textbf{Attention Mechanism.} The attention mechanism~\cite{46201} is shown in Algorithm~\ref{algo:attention}. In the summarization stage, Q, K, and V are matrices with the same dimension, while in the generation stage, Q is one single vector, and K, V are matrices. 

\begin{figure}[t]
    \centering
    \includegraphics[width=\columnwidth]{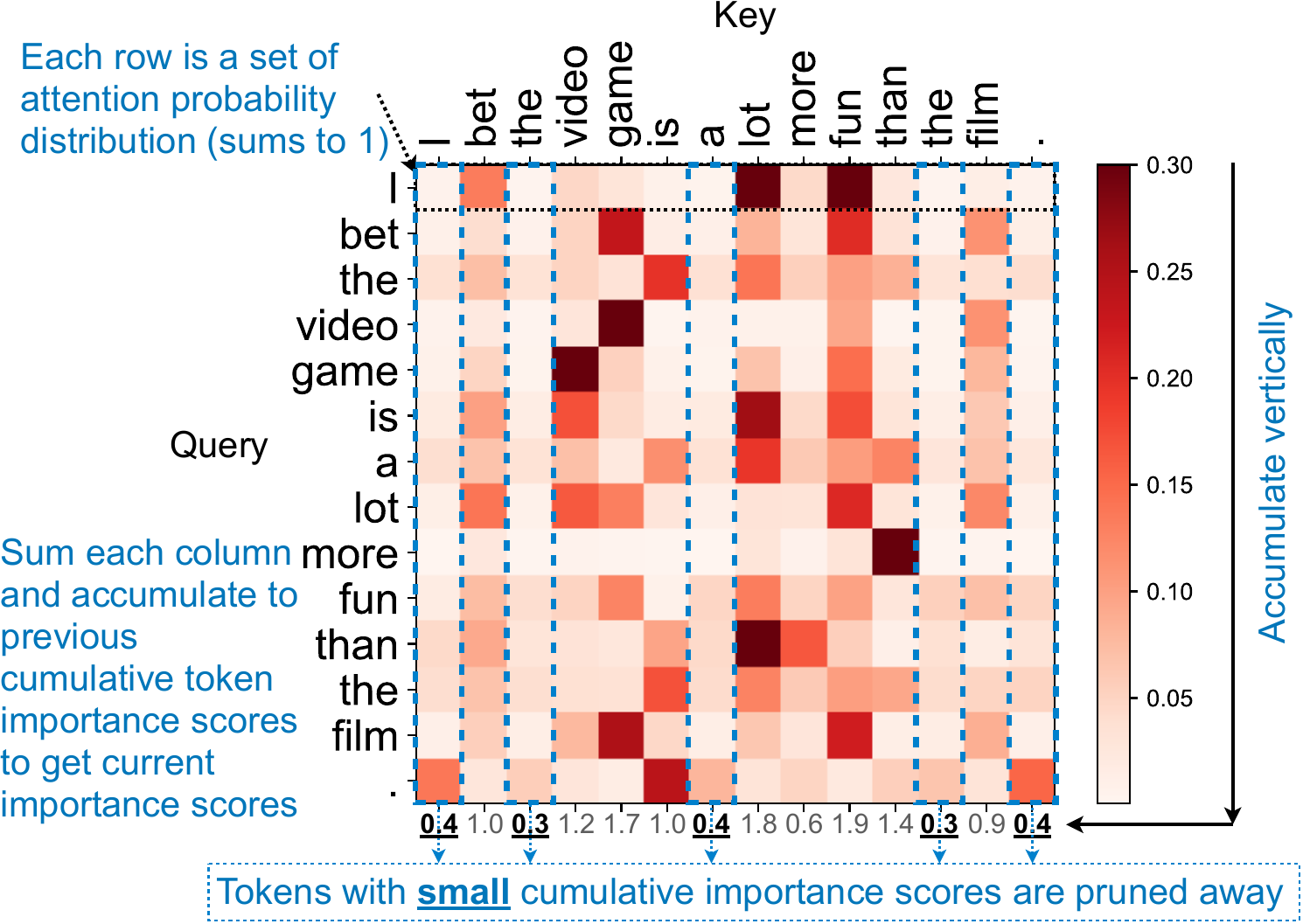}
    \vspace{-10pt}
    \caption{Attention probabilities for \bert are summed over each column to get importance scores. Tokens with small importance scores are pruned.}
    \vspace{-10pt}
    \label{fig:heatmap_bert}
\end{figure}

Attention has multiple \textit{heads}, each processing a chunk of Q, K, and V. Different heads capture various dependencies between tokens, some for long-term, some for short-term. Inside each head, $\mathrm{Q} \times \mathrm{K}^{T} / \mathrm{sqrt}(D)$ gives attention scores, which indicate whether two tokens are related. For instance in Figure~\ref{fig:heatmap_bert}, the score for `more' attending to `than' is large. After that, a row-wise softmax computes the attention probabilities. The exponential of softmax further enlarges the score for highly-related token pairs. The head's feature is then computed with $attention\_prob \times$V. This step lets each token fetch information from their cared tokens. Finally, multiple heads are concatenated together as the attention output. In \bert-Base and \gpttwo-Small, there are 12 blocks, and each attention layer has 12 heads. There are 24 blocks and 16 heads for each attention layer in \bert-Large and \gpttwo-Medium.

\subsection{Motivation}
\label{sec:motivation}
We profile the end-to-end latency of a \gpttwo model on multiple hardware platforms in Figure~\ref{fig:latency_breakdown}. Attention typically accounts for over 50\% latency, even though it only has around 10\% of overall \flops. In Figure~\ref{fig:latency_breakdown} right, around 73\% of the time is spent on data movements such as splitting heads, K and V concatenations, reshape, and transpose. 
GPUs and CPUs are well-optimized for matrix multiplications but are poor on the complex memory operations, thus slowing down attention. Therefore, it is necessary to build an accelerator to solve the attention bottleneck as a co-processor. The FC layers of NLP models are processed by GPUs, CPUs, or tensor algebra accelerators as they are highly-optimized for FC, and our \name co-processor handles all attention layers.

\section{Algorithmic Optimizations}
\label{sec:method}
\subsection{Cascade Token Pruning}

Plenty of unessential tokens exist in human languages, which can be pruned away to boost efficiency. 
Therefore, we propose cascade token pruning to assess token importance based on attention probabilities and remove trivial ones.
Cascade means that once a token is pruned, it is removed in all the following layers, so one layer only needs to process \emph{remaining} tokens from previous layers.

In cascade token pruning, tokens to be pruned are determined by an array of \emph{cumulative token importance scores}, one for each token (Figure~\ref{fig:methodology} and Algorithm~\ref{algo:cascade}). The scores are obtained by accumulating attention probabilities across multiple rounds of attention. The probability indicates whether a token is important to the sentence, because if the probability is large, then the outputs are more influenced by the corresponding token. Specifically, $attention\_out = attention\_prob \times$V while V is computed from input features $block\_in$ (see Figure~\ref{fig:background}). The $block\_in$ of the first block are directly from input tokens. For latter blocks, V vectors are still largely determined by input tokens because of residual connections. Therefore, the probability is an indicator of the token importance, and the accumulations across several heads and layers make the importance more reliable. For example, in Figure~$\ref{fig:heatmap_bert}$, many tokens attend to the word `fun', implying its high usefulness. In each head, the scores are accumulated by $L_0$ (number of query vectors) times in the summarization stage and one time in the generation stage. In \bert, we accumulate importance scores in former heads and layers and apply token pruning to latter ones. In \gpttwo, we further accumulate importance scores \emph{across generation iterations} because intuitively, the unimportant tokens for one token generation should also be unimportant to others.
With the cumulative importance scores, we can remove a pre-defined pruning ratio of tokens. Once a token is pruned, the QKV of it will \emph{never} be used in all the following attention heads and layers; in every layer/head, several new tokens can be selected and pruned away, thus being \emph{global and cascade}. 
Token pruning can reduce the computation and memory access of both attention, and also FC layers outside attention. On eight \gpttwo benchmarks, token pruning can achieve \prunegpttwodramsaving\x reduction of DRAM access.

\subsection{Cascade Head Pruning}
Each QKV vector has multiple chunks corresponding to multiple heads, which are used to capture various token dependency relationships. However, some of the heads are redundant~\cite{voita2019analyzing} and have little influence on outputs. 
Token pruning reduces the sentence length. Head pruning reduces the feature length. Hence, redundancies in both dimensions are removed.
The head importance is calculated by accumulating the absolute value of $attention\_out$ elements of each head across layers to get \emph{cumulative head importance scores} (Figure~\ref{fig:methodology} and Algorithm~\ref{algo:cascade}). The magnitude of the head's outputs indicates its importance because there exists one FC layer processing the concatenation of all heads. If the magnitude of one head is large, the outputs of the FC layer and the whole block $block\_out$ will be more heavily influenced by the head. Similar to token pruning, head pruning is also cascaded: once removed, a head will not appear in the following layers.

\begin{figure}[t]
    \centering
    \includegraphics[width=1\columnwidth]{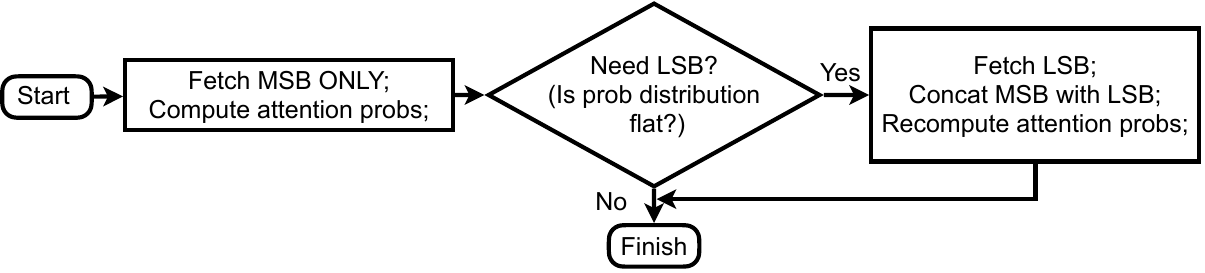}
    \vspace{-20pt}
    \caption{Progressive Quantization. MSBs are fetched first. Only if the resulting probability distribution is flat, LSBs are fetched, and attention probabilities are recomputed -- this trades computation for memory saving. 
    }
    \vspace{-10pt}
    \label{fig:pq_flow}
\end{figure}

\subsection{Local Value Pruning}
\name also supports local Value (V) pruning, which is performed after \softmax. the V vectors to be pruned are decided \emph{solely with the current head's attention probabilities.} A pre-defined ratio of V vectors with the smallest attention probabilities are pruned and will not be fetched for the $attention\_prob \times$V computation.
Compared to cascade token pruning which removes Q, K, and V of pruned tokens for current and all following attention heads and layers, local V pruning only removes V vectors of the current head.

\subsection{Progressive Quantization}
\label{sec:method:dyna}
\softmax layers are abundant in attention, which allows us to apply more aggressive \lowprec than CNN/RNN models because \softmax can reduce quantization error. Softmax for attention probabilities is:
$p_{i} = e^{s_{i}} / \sum_{j=0}^{L_1-1} e^{s_{j}} $, 
where $L_1$ is the number of K vectors, $p$ is attention probability, and $s$ is attention score. Quantization on Q and K can be considered as adding a small error $\Delta s$ to the attention score $s$. We examine the influence of $\Delta s$ on output attention probabilities by computing the softmax derivative:
\begin{equation}
\label{equ:dsoftmax}
\begin{aligned}
    \mathrm{If \ } i = j: \frac{\partial{p_i}}{\partial{s_j}}
    & = \frac{\partial{\frac{e^{s_{i}}}{\sum_{i=0}^{L_1-1} e^{s_{i}}}}}{\partial{s_i}}
    = p_i \cdot (1-p_i) \\
    \mathrm{If \ } i \neq j: \frac{\partial{p_i}}{\partial{s_j}}
    & = \frac{\partial{\frac{e^{s_{i}}}{\sum_{i=0}^{L_1-1} e^{s_{i}}}}}{\partial{s_j}} 
    = -p_i \cdot p_j
\end{aligned}
\end{equation}
 Without loss of generality, we assume $s_0$ changes by $\Delta s_0 > 0$, and sum the absolute errors of all output with Equation~\ref{equ:dsoftmax}:
\begin{equation}
\label{equ:error}
\begin{aligned}
error 
& = \lvert \Delta s_0 \cdot p_0 \cdot(1-p_0)  \rvert + \sum_{i=1}^{L_1-1} \lvert - \Delta s_0 \cdot p_0 \cdot p_i \rvert \\
& = \Delta s_0 \cdot (2\cdot p_0 \cdot (1-p_0)) < \Delta s_0
\end{aligned}
\end{equation}
Since $0 \le p_0 \le 1$, $2 p_0  (1-p_0)$ is always smaller than 0.5, so the total quantization error is reduced after \softmax.

\begin{figure}[t]
    \centering
    \includegraphics[width=1\columnwidth]{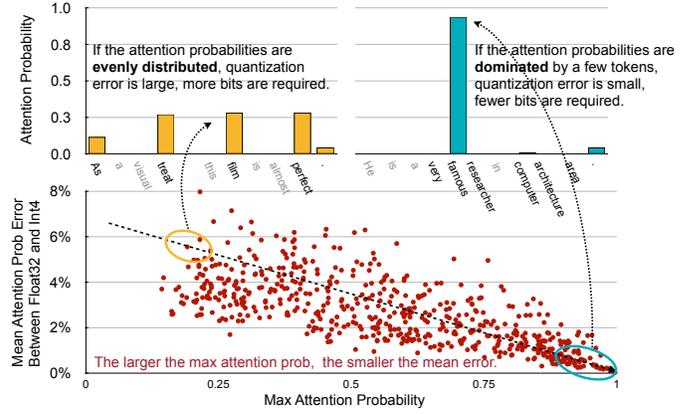}
    \vspace{-20pt}
    \caption{When the attention probability is evenly distributed (left), the quantization error is large; both MSB and LSB are required. When dominant probability exists (right), the quantization error is small; only MSB is required. Grey tokens are pruned by token pruning, thus no need to compute their attention probabilities.
}
    \vspace{-10pt}
    \label{fig:dynamic_precision}
\end{figure}

\begin{figure*}[t]
    \centering
    \includegraphics[width=1\textwidth]{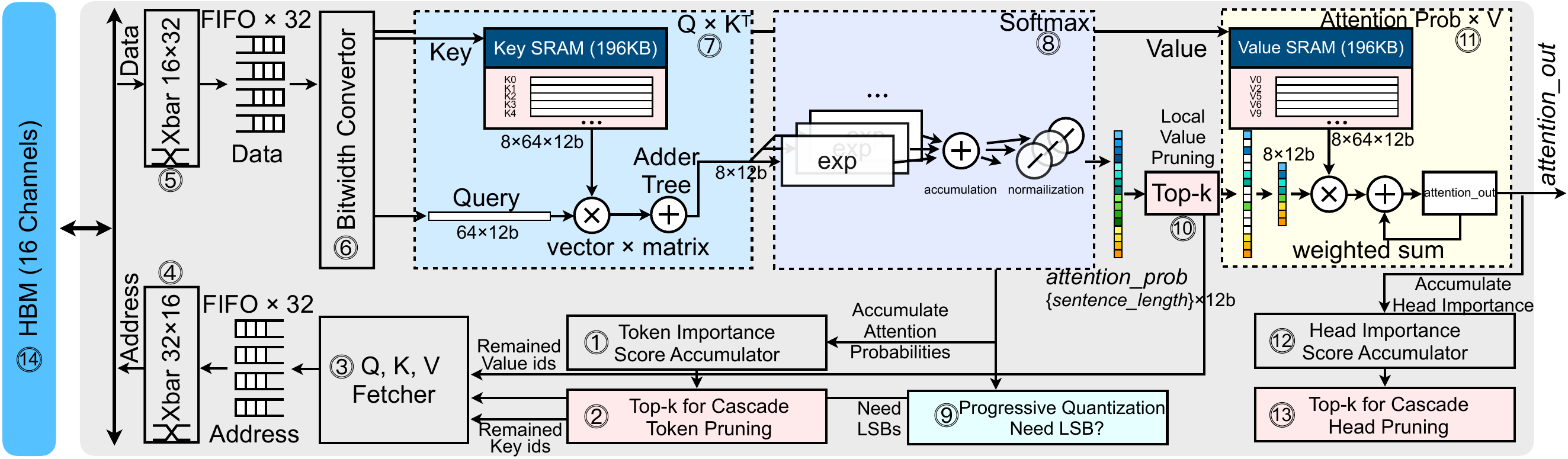}
    \vspace{-20pt}
    \caption{SpAtten Architecture Overview. Modules on the critical path (6,7,8,10,11) are fully pipelined to maximize the throughput.}
    \vspace{-10pt}
    \label{fig:arch_overview}
\end{figure*}

On top of static \lowprec, we propose \emph{progressive \lowprec} (Figure~\ref{fig:pq_flow}) to progressively increase the input \bitwidth if aggressive quantization hurts accuracy. An interesting phenomenon in Figure~\ref{fig:dynamic_precision} shows that if the attention probability distribution is dominated by a few tokens, then the 4-bit quantization error is smaller; if flat, the error is larger. Intuitively, dominant tokens are semantically important; thus cannot be easily influenced by quantization. Theoretically, errors are proportional to $p(1-p)$ (Equation~\ref{equ:error}). When probability dominators exist, $p$ is closer to zero or one; thus errors are smaller. Otherwise, errors are larger.
Therefore, for robust layers with probability dominators, the \bitwidth can be small; while other sensitive layers should have more bits. In Figure~\ref{fig:pq_flow}, we firstly apply an aggressive \bitwidth (only fetch MSBs) for inputs and compute the attention probabilities. If the max of computed probability is smaller than a threshold, indicating the distribution is flat, we will fetch LSBs for inputs and recompute the attention probabilities for once. Otherwise, LSBs are not needed. Intuitively, progressive quantization finds which input sample is \emph{more difficult} and applies a higher bitwidth instead of using a high bitwidth universally, thus reducing DRAM access under the same accuracy.

For fast interactions between \name and hardware for FC parts, we conduct \emph{linear symmetric quantization}, which is much faster than K-Means quantization. We have five different MSB+LSB settings: 4+4, 6+4, 8+4, 10+4, and 12+4. The settings can be different across tasks but are the same within one task. Different inputs of one task determine \emph{whether to fetch LSB on the fly}. We store MSBs continuously and LSBs continuously in DRAM, so that they can be fetched separately.
Progressive \lowprec trades more computation for less memory access and can effectively accelerate memory-bounded generative models such as \gpttwo. It also improves energy efficiency since DRAM access takes around 70\% of \name power, much expensive than computation.  
On average, only 5.9\% input samples require LSB. For \bert, we only apply static quantization because \bert models are computation-bounded, and fetching LSB for recomputation will degrade \bert's performance.

\section{Hardware Architecture}
\label{sec:arch}
\subsection{Overview}
\begin{figure}[t]
    \centering
    \includegraphics[width=\columnwidth]{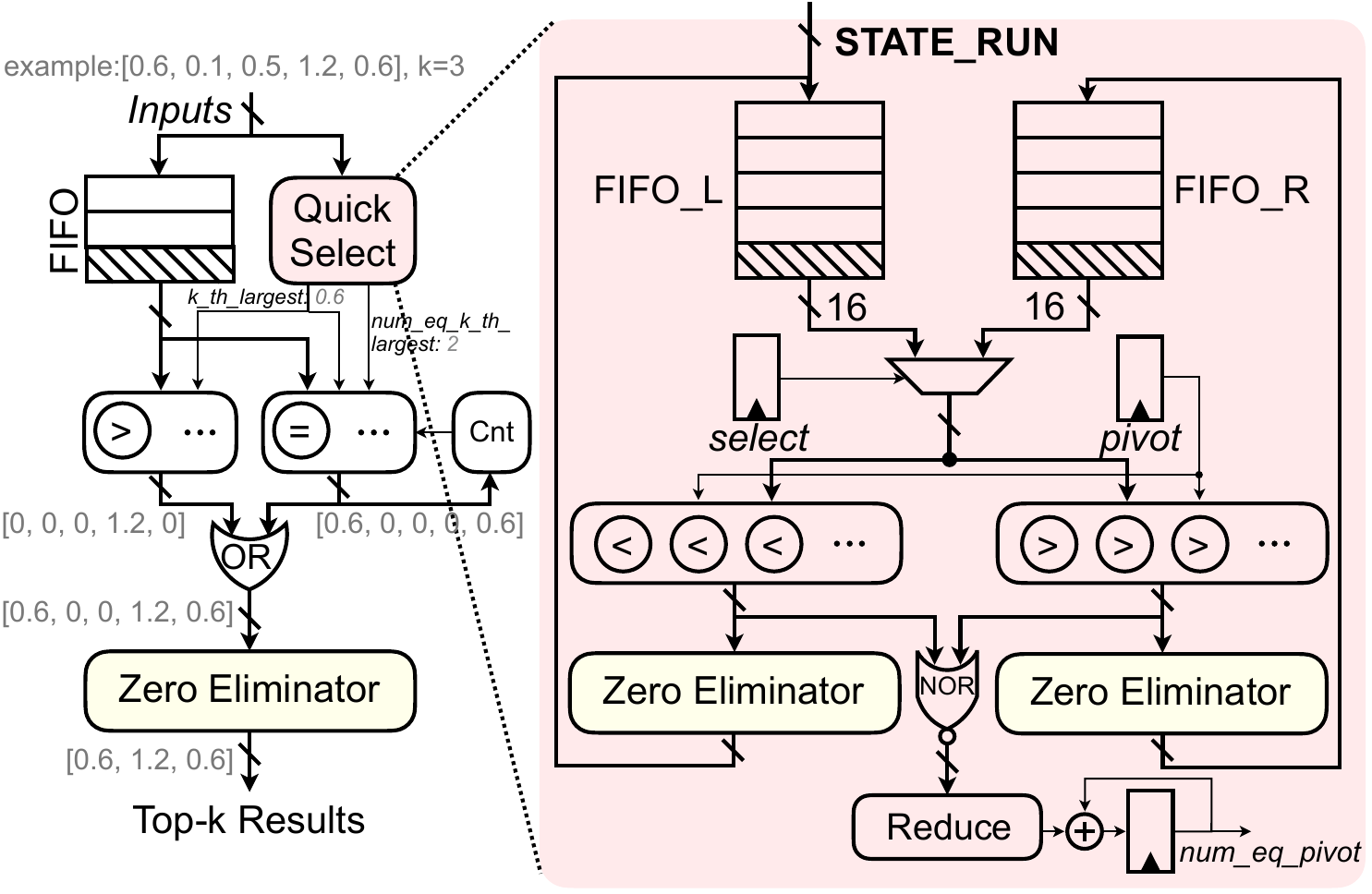}
     \vspace{-10pt}
    \caption{High Parallelism \topk Engine.
    }
     \vspace{-10pt}
    \label{fig:topk}
\end{figure}

An overview of \spatten is shown in Figure~\ref{fig:arch_overview}. 
To support token/head pruning, a novel \topk engine (Figure~\ref{fig:topk}) is designed to rank the token/head importance. Pruning reduces computation and memory traffic but incurs random access. Thus, a crossbar is applied to process the addresses, keep each memory channel busy, and increase the bandwidth utilization rate. To support progressive \lowprec, we implement an on-chip \bitwidth converter to handle the splits of fetched bits and concatenations of MSBs and LSBs.

\spatten processes attention head by head and query by query, thus well balancing the pruning granularity and parallelism. One query of a head is fed to the pipeline at a time, enabling token pruning in both head-wise and layer-wise granularity. Inner-head parallelism can keep all on-chip computation resources busy, so no need for inter-head parallelism. In the summarization stage, K and V that survive cascade token pruning are fetched to the on-chip SRAM and will be \emph{reused} across multiple queries. In the generation stage, the Q is a single vector, so there is no reuse of K and V, and no need to store them in the on-chip SRAM.

\begin{figure}[t]
    \centering
    \includegraphics[width=1\columnwidth]{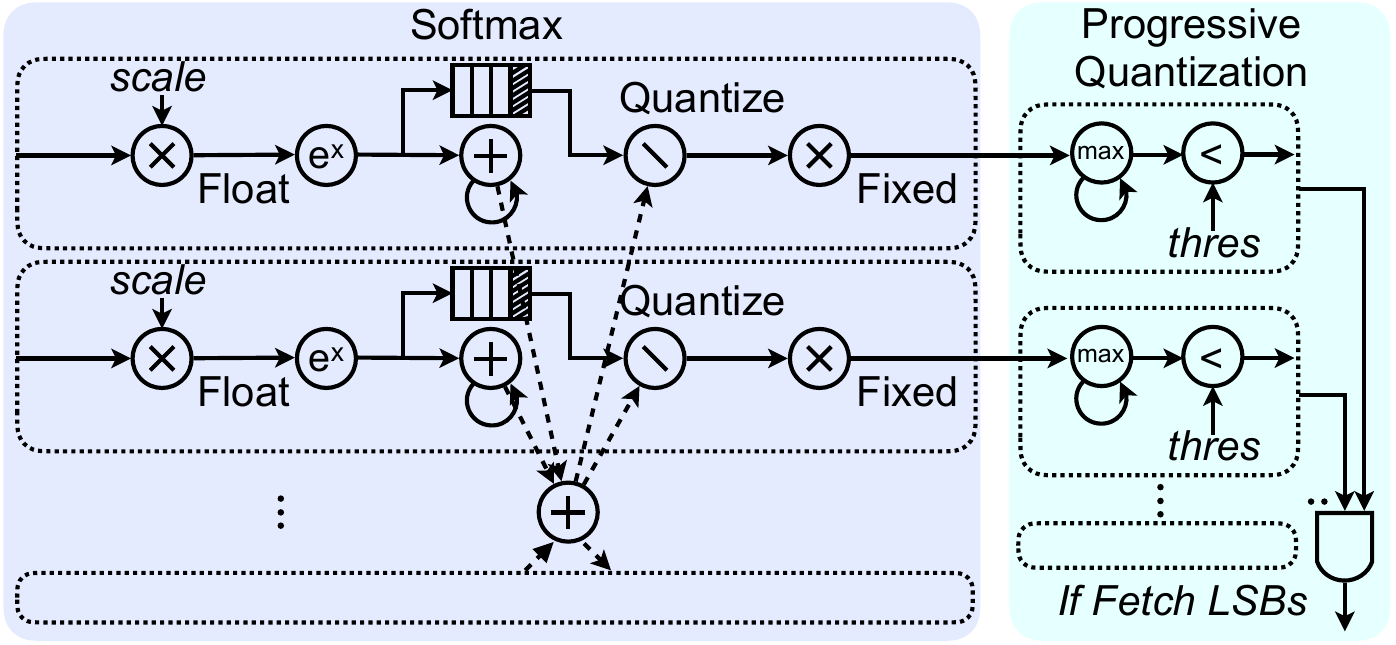}
    \vspace{-20pt}
    \caption{Softmax and progressive quantization determination modules.}
    \vspace{-10pt}
    \label{fig:softmax}
\end{figure}

For each fetched Q, the \topk engine (Figure~\ref{fig:topk}) first ranks the token importance scores and get $k$ most important Ks. A data fetcher then computes the Ks' addresses and feeds them to a 32$\times$16 crossbar for 16 HBM channels. It then gets data back through a reverse 16$\times$32 crossbar to preserve the correct order. 
Q and K are processed by a matrix-vector multiplication module (Figure~\ref{fig:qxk}) to get the attention scores. 
A Softmax module (Figure~\ref{fig:softmax} left) then processes the attention scores to get attention probabilities, and sends them to the progressive quantization module (Figure~\ref{fig:softmax} right) to decide whether LSBs are required. The probabilities are also sent to the token importance score accumulator to perform accumulations.
After that, the local Value pruning \topk engine gets the probabilities, computes $k$ most locally important Vs, and sends their indices to the data fetcher. Finally, the remaining probabilities are multiplied with fetched V, getting attention outputs. After computing one head, the head importance score will be accumulated. After finishing all heads in a layer, a top-k module prunes unimportant heads, which will not be computed in any following layers. 

Our dataflow guarantees to avoid fetching pruned tokens and value vectors, thus bringing DRAM access reductions. The critical path (module 6,7,8,10,11) is fully pipelined. The accumulation of token/head importance scores and token/head top-k are performed \emph{in parallel} with the critical path. For module 3 branches, when multiple sources of Q/K/V requests come simultaneously, the fetcher processes requests one by one and sends addresses to FIFOs. The branches after module 8/11 send $attention\_prob/attention\_out$ to accumulators and the next corresponding computation module simultaneously. For the module 9 branch, if LSB is required, it discards $attention\_prob$ and initiates recomputation; then modules 10 and 11 will be idle, waiting for recomputed attention probabilities. For module 10 branch, fetching un-pruned V from DRAM is part of the coarse-grained pipeline.

The on-chip memory system has two main SRAMs for Key and Value, 196KB each in module 7 and module 11. They store K and V from QKV Fetcher. Since we process queries one by one, the Q vector is stored in registers. We also have 32 64-depth$\times$8B address FIFOs after QKV fetcher and 32 64-depth$\times$16B data FIFOs before bitwidth converter.

\begin{figure}[t]
    \centering
    \includegraphics[width=\columnwidth]{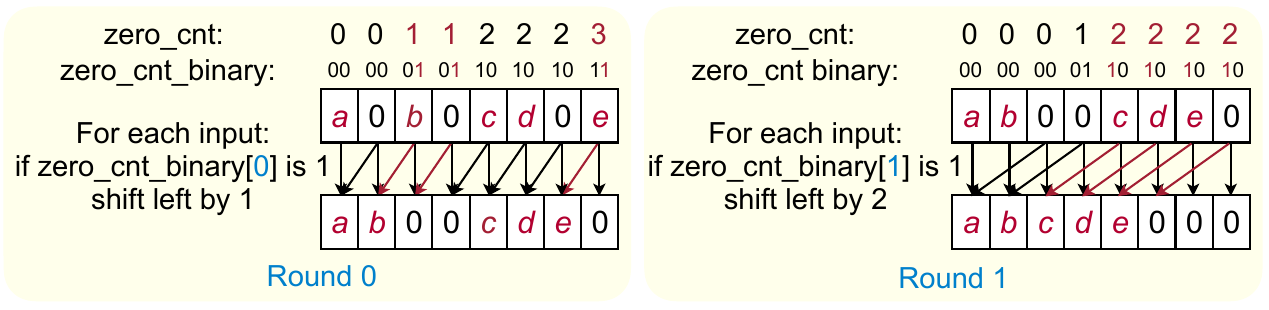}
     \vspace{-20pt}
    \caption{Zero Eliminator. A zero eliminator of $n$ elements has $\log n$ stages.}
     \vspace{-20pt}
    \label{fig:ze}
\end{figure}

\begin{algorithm}[!t]
\footnotesize{
\SetKwInOut{Input}{Input}
    \textbf{Input:} Top-k $k$, Data vector $inputs$; \\
    FIFO\_L, FIFO\_R depth: 64; FIFO $=$ [FIFO\_L, FIFO\_R]; \\
    Initialize FIFO\_L with $inputs$, FIFO\_R $= \phi$; \\
    $target = k$, $num\_eq\_pivot = 0$;\\
    \SetAlgoLined
    \textbf{START:} \\
    \uIf{ $\mathrm{size(FIFO\_R)} + num\_eq\_pivot \le target$ }{
    \tcc{ the selected pivot is too large}
    $target = target - \mathrm{size(FIFO\_R)}-num\_eq\_pivot$; \\
    FIFO\_R $ =\phi$; $select = 0$; $pivot = $ FIFO\_L[rand]; \\
    Goto \textbf{STATE\_RUN};\\
    
    }
    
    \uElseIf{$\mathrm{size(FIFO\_R)} > target$}{
    \tcc{ the selected pivot is too small}
    FIFO\_L $=\phi$; $select = 1$; $pivot = $ FIFO\_R[rand]; \\
    Goto \textbf{STATE\_RUN}; \\
    }
    
    \Else{ \tcc{
    $\mathrm{size(FIFO\_R)} \le target$
    $\mathrm{size(FIFO\_R)} + num\_eq\_pivot > target $
    }
    $k\_th\_largest = pivot$; \\ $num\_eq\_k\_th\_largest = target$-size(FIFO\_R); \\
    \textbf{Output:} [$k\_th\_largest, num\_eq\_k\_th\_largest$]; \\
    }
    
    \colorbox{topkcolor}{\vbox{
    \textbf{STATE\_RUN:} (Figure \ref{fig:topk} Right) \\
    \tcc{items smaller than pivot $\rightarrow$ FIFO\_L \\
    items larger than pivot $\rightarrow$ FIFO\_R}
     $num\_eq\_pivot = 0$; \\
    \For{$i = 0 \mathbf{ \ to \ } \mathrm{size(FIFO} [select])-1$}{
    Pop FIFO[$select$] as $item$;\\
    \uIf{$item < pivot$}{
    Push $item$ to FIFO\_L;\\
    }
    
    \uElseIf{$item > pivot$}{Push $item$ to FIFO\_R;\\}
    \Else{ \tcc{$item == pivot$}
    $num\_eq\_pivot = num\_eq\_pivot + 1$;\\}
    }
    Goto \textbf{START};
    }}

    \caption{\topk Engine}
    }
    \label{algorithm:topk}

\afterpage{\global\setlength{\textfloatsep}{\oldtextfloatsep}}

\end{algorithm}

\subsection{Top-k Engine}
To support cascade token/head pruning and local value pruning, we need to find the top $k$ elements of an array. A na\"ive solution is to use a sorter to sort the original array and directly output the first $k$ elements. However, it needs $O(n\cdot \log n)$ time complexity and $O(n\cdot \log^2n)$ space sorting network, and also completely randomizes data fetching.
Instead, we design a novel \topk engine (Figure~\ref{fig:topk}) with a quick-select module to find the $k^{\mathrm{th}}$ largest element as a threshold to filter the input array. It has much lower time complexity ($O(n)$ on average) and keeps the original order of inputs.
The engine leverages a randomly chosen pivot to partition the input array into two parts: elements smaller or larger than the pivot. It has two FIFOs, FIFO\_L and FIFO\_R, to store the partitioned arrays. An array to be partitioned will be fed into two comparator arrays and compared with the pivot. The left/right comparator array only preserves elements smaller/larger than the pivot. Others will be set to zeros and eliminated by a zero eliminator. Quick-select runs iteratively until the $k^{\mathrm{th}}$ largest element is found. The control logic is in Algorithm~\ref{algorithm:topk}.
Afterwards, the $k^{\mathrm{th}}$ largest element is used to filter the input array, which is buffered in another FIFO (Figure~\ref{fig:topk} left) before the quick-select process. The filtered outputs will be processed by another zero eliminator to finally get the top-k elements.
It can be easily scaled up with more comparators. In \spatten, we apply 16 comparators in each array to make it not the whole pipeline's bottleneck. Since the frequency of head pruning is much lower than token pruning and local Value pruning, we reuse the token pruning \topk engine for head pruning. Therefore, we have two \topk engines in \name architecture.

\subsection{Zero Eliminator}
We design an innovative logic for the zero eliminator, which first uses a prefix sum module to calculate the number of zeros before each element $zero\_cnt$. These $zero\_cnt$ will guide a logN layer shifter to shift the input array by 1, 2, 4, ... positions. In the $n^{\mathrm{th}}$ layer, whether to shift an element is determined by the $n^{\mathrm{th}}$ bit of its corresponding $zero\_cnt$. An example is illustrated in Figure~\ref{fig:ze}.

The top-k engine equipped with zero-eliminators has much higher parallelism than a direct implementation of quick select. Without high parallelism, the performance will be bottlenecked by finding top-k. In Figure~\ref{fig:perfbreak}, we show that SpAtten with a parallelized top-k engine can achieve 3$\times$ speedup over a baseline top-k engine with parallelism=1. We also compare a regular full sorting unit (a Batcher's Odd-Even Sorter~\cite{10.5555/280635} to perform merge-sort) to the worst case of the top-k engine (selecting the median) with an input length of 1024. Experimental results show that we can achieve 1.4\x higher throughput with 3.5\x smaller power consumption over the full sorting unit.

\begin{figure}[t]
    \centering
    \includegraphics[width=\columnwidth]{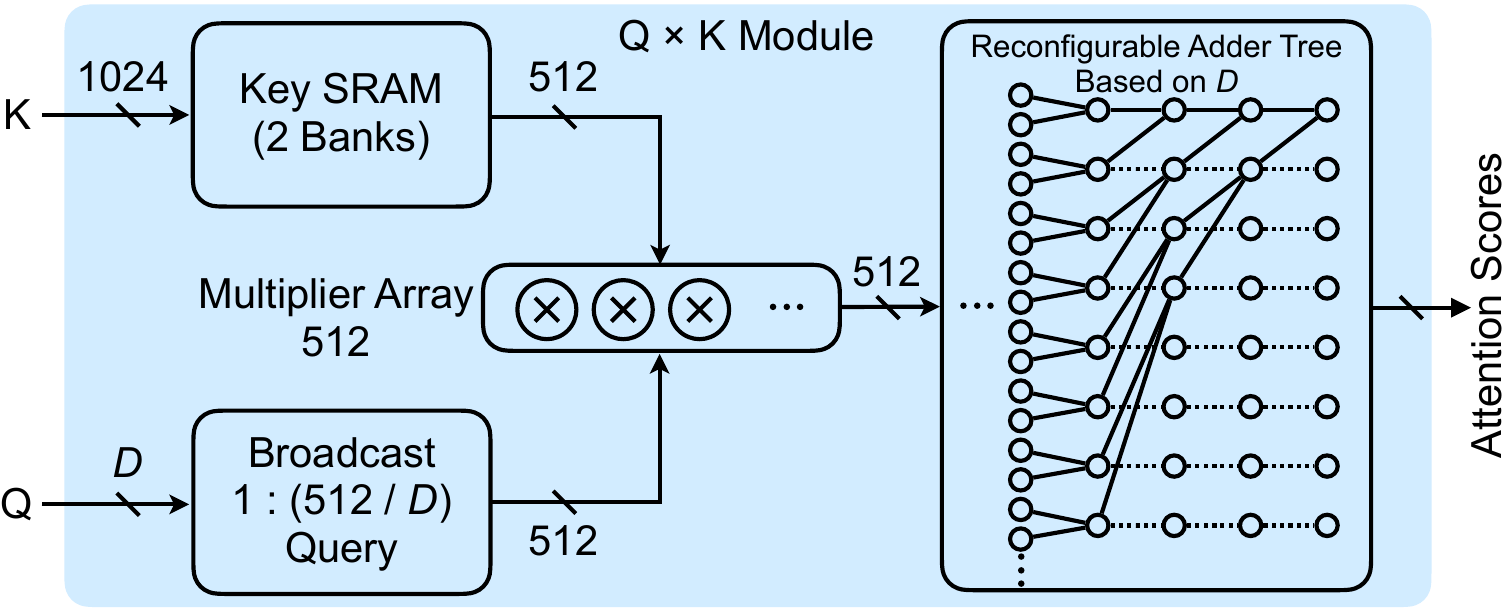}
    \vspace{-15pt}
    \caption{Query-Key multiplication module with a reconfigurable adder tree. The adder tree can output a various number of partial sums according to the query vector's dimension.}
    \vspace{-15pt}
    \label{fig:qxk}
\end{figure}

\subsection{Data Fetcher and Bitwidth Converter}
The Q-K-V data fetcher is designed to send multiple random read requests per cycle to all 16 HBM channels, where the QKV are interleaved in different channels. We use a 32-to-16 crossbar to route these read requests to the correct channels. The master side is larger than the slave side. There is no memory access conflict because the crossbar generates at most one memory request for each channel at a time.

In order to support progressive quantization but avoid complex logic overheads, we enforce on-chip SRAMs and multipliers to have a fixed bitwidth. We use a bitwidth converter to convert the data loaded from DRAM (4,8,12 bits) uniformly into on-chip bitwidth (12 bits). The converter consists of MUXes to select correct bits from the input and a shifter to allow reading data from an unaligned address.

\subsection{Query-Key Multiplication Module}
The query-key multiplication module (Figure~\ref{fig:qxk}) is designed to calculate the matrix-vector multiplication between K and Q. In each cycle, a row of the key matrix K$_i$ is loaded from the K SRAM, multiplied by Q with a multiplier array, and fed to an adder tree. Adder tree then computes attention scores by reducing all multiplied results $s_i = \sum_j \mathrm{K}_{ij} \times \mathrm{Q}_j$.
We use 512 multipliers in this module to fully utilize the DRAM bandwidth. To support queries and keys with dimension $D$ lower than 512, we get 512/$D$ attention scores in each cycle. The corresponding multiple K$_i$s are packed into the same line in the Key SRAM, and the query is broadcast for 512/$D$ times so that all K$_i$s can have access to the same Q. The adder tree is also configurable to output the results of the last several layers, making it function like 512/$D$ separate D-way adder trees capable of producing multiple results $s_i$.

\subsection{Softmax and Progressive Quantization}
The fixed-point attention scores $s$ from query-key multiplication are first dequantized using a scaling factor. The attention score normalization factor sqrt($D$) is also included in the scaling factor, so that we can perform attention score dequantization and normalization simultaneously. After that, a pipeline of floating-point exponential, accumulation, and division operations are applied to calculate the Softmax results $e^{s_i}/\sum_j e^{s_j}$. The results are finally quantized again so that the operations after Softmax can be performed in fixed-point.

The Softmax results are then fed to the progressive quantization determination module to examine whether LSBs are required. Specifically, we compare the max attention probability with a pre-defined threshold. If smaller than the threshold, the Q-K-V data fetcher will be informed to fetch the LSBs.

\subsection{Attention Prob-Value Multiplication}
The attention prob-value multiplication unit takes the attention probabilities as inputs, multiplies them with the Vs and then accumulates to get attention outputs $A_j = \sum_i attention\_prob_i \times \mathrm{V}_{ij}$. It contains another broadcast-multiply-reduce pipeline, similar to the one in the query-key multiplication module, to support the processing of multiple attention probabilities at the same time. The module has 512 multipliers. The attention probabilities are broadcast for $D$ times, and the adder tree is configured to function like $D$ 512/$D$-way adder trees. The $attention\_out$ are in 12 bits.

\begin{table}[t]
\renewcommand*{\arraystretch}{0.5}
\setlength{\tabcolsep}{8pt}
\footnotesize
\centering
\caption{Architectural Setups of \spattenfull.}
\vspace{-5pt}
\begin{tabular}{l|l}
\toprule
  \multirow{2}{*}{Q-K-V Fetcher} & A 32$\times$16 address crossbar and a 16$\times$32 \\
   & data crossbar; each port has a 64-depth FIFO. \\ \midrule
 \multirow{2}{*}{Q $\times$ K} & 196KB Key SRAM; 512$\times$12-bit multipliers; \\
 & Adder tree outputs at most 8 items/cycle. \\ \midrule
 Softmax & FIFO depth for softmax: 128; Parallelism: 8. \\ \midrule
 Attention\_Prob $\times$ V & 196KB Value SRAM; 512$\times$12-bit multipliers.  \\ \midrule
 \multirow{3}{*}{HBM} & HBM2, 16$\times$128-bit HBM channels @ 2GHz;\\
 & each channel has 2$\times$64-bit pseudo-channels \\
 & and provides 32GB/s bandwidth. \\
 \bottomrule
\end{tabular}
\vspace{-10pt}
\label{tab:setup}
\end{table}

\begin{table}[t]
\centering
\caption{Power Breakdown of \spattenfull.}
\vspace{-5pt}
\setlength{\tabcolsep}{9pt}
\footnotesize
\centering
\begin{tabular}{l|c|c|c|c}
\toprule
& Computation Logic & SRAM & DRAM & Overall \\
\midrule
Power & \powercompute W & \powersram W & \powerdram W & \powerspattenfull W \\
\bottomrule

\end{tabular}
\label{tab:powerbreak}
\vspace{-10pt}
\end{table}

\section{Evaluation}
\subsection{Evaluation Methodology}
\label{sec:evalmethod}
We implement \name with SpinalHDL and compiled to RTL, and simulate each application using Verilator to get the cycle numbers. For HBM modeling, we use Ramulator~\cite{kim2015ramulator} with HBM2 settings. We synthesize two versions of SpAtten: {\spattenfull} and \spattensmall. {\spattensmall} is only used for fair comparisons with \mnnfast and \athree. The parameters for {\spattenfull} are listed in Table~\ref{tab:setup}. The scale of {\spattensmall} is 1/8 of {\spattenfull} and contains 128 multipliers.
We synthesize \name using Cadence Genus under TSMC 40nm library to estimate the area and power consumption of the logic, including all fixed-point adders and multipliers. We get the number of floating-point operations in \softmax from the simulator. The exponential function is approximated with Taylor expansion to the 5$^{\mathrm{th}}$ order~\cite{7004740} and performed with floating multiplication accumulation units (FMA). The power and area of FMA are obtained from~\cite{salehi2015energy}. We perform the division and estimate power and area with the floating-point unit (FPU) from~\cite{salehi2015energy}. The FMAs and FPUs are in 45nm technology, and we use them as an upper bound estimation of 40nm units. 
We also obtain the width, size, and the number of read/write of each SRAM and FIFO from the simulator and use CACTI~\cite{cacti} to estimate the energy and area of SRAMs and FIFOs. For HBM, we simulate the number of row activation, read/write with Ramulator, and use the energy numbers from \cite{o2017fine} to calculate overall energy. 

We extensively select various hardware platforms as the evaluation baselines, including server GPU (NVIDIA \titanxp), mobile GPU (NVIDIA Jetson Nano), server CPU (Intel Xeon E5-2640 v4 @ 2.40GHz), mobile CPU (4-core ARM A53 CPU on a Raspberry Pi-4), and state-of-the-art accelerators \athree~\cite{ham20203} and \mnnfast~\cite{8980322}. For GPUs and CPUs, we run attention with PyTorch and use cuDNN on GPU and MKL on CPU, which are well-optimized libraries.
\texttt{torch.cuda.Event} on GPU, and \texttt{time.time} on CPU are used to measure the latencies. We measure the power with \texttt{nvidia-smi} and \texttt{pcm-power} for \titanxp \ and \xeon, respectively. For \nano \xspace and \rasp, we use a power meter to get power. For latency measurements, we repeat 1000 times, remove the largest 15\% and smallest 15\%, and average the remaining. For power measurements, we first measure the system's idle power, and then repeatedly run workloads and get the total power. The dynamic power is total power minus idle power. 

We evaluate \name on attention layers of two discriminative models: \bert-Base, \bert-Large, and two generative models: \gpttwo-Small and \gpttwo-Medium. Tasks for \bert are nine from  GLUE~\cite{wang-etal-2018-glue}, V1.1 and V2.0 of SQuAD~\cite{rajpurkar2016squad}; For \gpttwo, we use language modeling task on four datasets: Wikitext-2~\cite{merity2016pointer}, Wikitext-103~\cite{merity2016pointer}, Pen Tree Bank~\cite{marcus-etal-1993-building} and One-Billion Word~\cite{chelba2013one}. In total, we have \numbenchmark benchmarks. 
For all tasks, finetuning is performed for 2 hours on average on GPU after token pruning to recover accuracy.
For each task, we try multiple sets of token/head pruning ratios and quantization \bitwidths to not lose accuracy, except 2\% for BERT-large on \squad tasks. Given the same overall pruning ratio, ratios among layers/heads do not have a significant influence. We typically keep the 15\% front layers un-pruned, then compute the average ratio of the rest layers $r_{avg}$. We set a start ratio $r_{start}$ and an end ratio $r_{end}$, $r_{start} + r_{end} = 2
\times r_{avg}$ and interpolate the ratios of the rest layers. For head pruning, we keep 30\% front layers un-pruned and apply a similar method as token pruning. For progressive quantization, the typical max attention probability threshold is 0.1, and the common MSB+LSB combinations are 6+4 and 8+4.

To measure \bert latency, we set input sentence length as the average length of the each task's dev set. For \gpttwo models, we set the initial length of the input sentence as 992 and measure the latency of generating 32 tokens. The energy efficiency of the models is assessed by energy consumption, which is power$\times$latency.

\subsection{Experimental Results}
\begin{figure}[t]
    \centering
    \includegraphics[width=1\columnwidth]{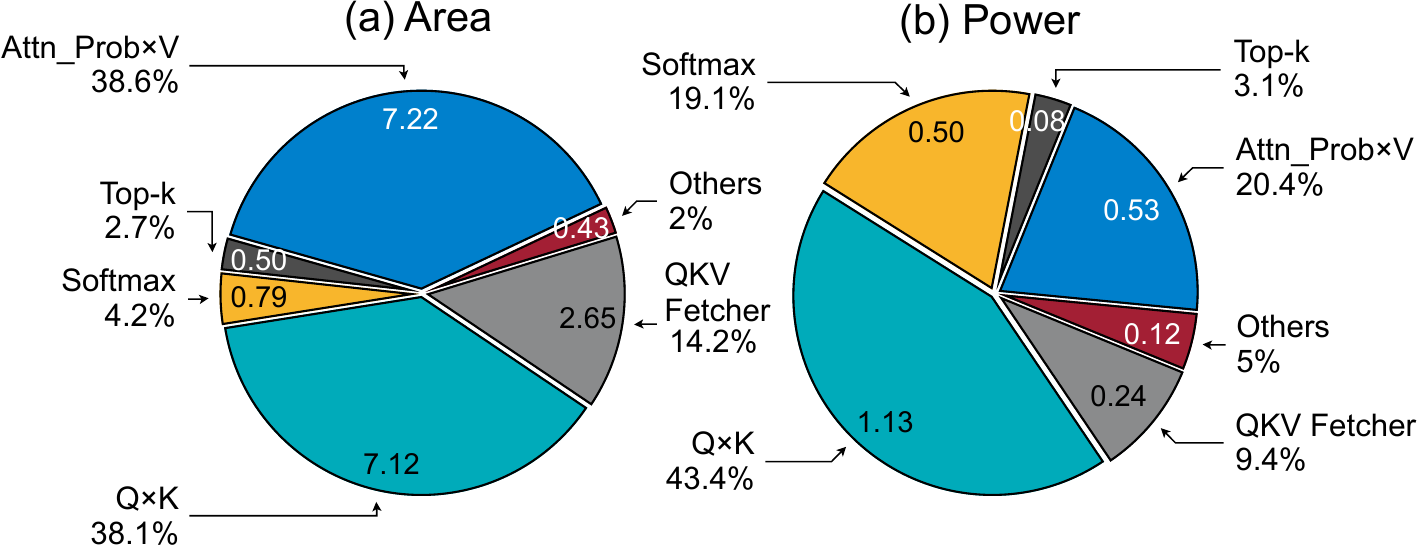}
    \caption{On-chip (a) Area and (b) Power Breakdowns of \spattenfull. 
    }
    \vspace{-15pt}
    \label{fig:pie}
\end{figure}
\begin{figure*}[t]
    \centering
    \includegraphics[width=1\textwidth]{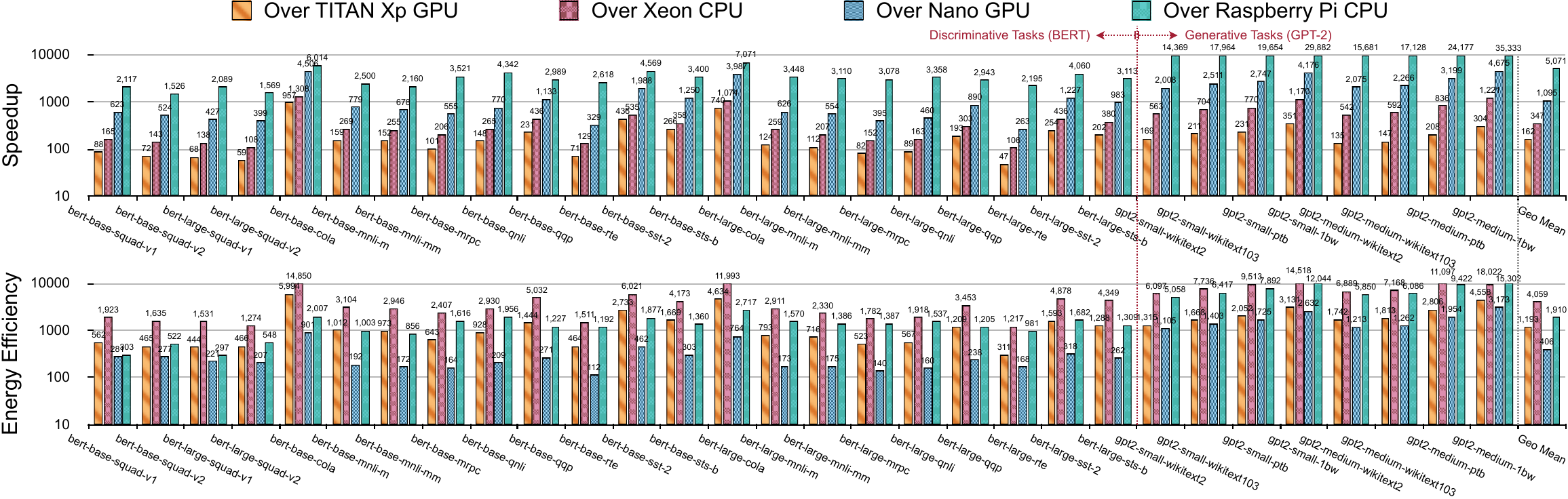}
    \vspace{-10pt}
    \caption{Speedup and Energy Efficiency of {\spattenfull} over baselines on attention layers. \name can accelerate both discriminative and generative models. 
    }
    \vspace{-10pt}
    \label{fig:perfcomp}
\end{figure*}

\textbf{Throughput, Power, and Area.}
\name prunes tokens and value vectors with cascade token pruning and local value pruning by \prunedramsaving\x (all models average), and \prunegpttwodramsaving\x (\gpttwo models average). Cascade head pruning has \hpdramsaving\x reduction on average. Note that the pruning ratio can be larger when the input sentence of a task is longer because they contain more redundancy. GPT-2 models have longer inputs than BERT, so their pruning ratios can be larger.
\name reduces the computation by \totalcompsaving\x and DRAM access by \totaldramsaving\x on average. It achieves \spattenbertflops\tflopss on 22 computation-bounded \bert models and \spattengpttwoflops\tflopss on 8 memory-bounded \gpttwo models. {\spattenfull} consumes \powerspattenfull W power as in Table~\ref{tab:powerbreak} breakdown and is \areaspattenfull mm$^2$ in area. 
Figure~\ref{fig:pie} shows the area and on-chip power breakdown. 
The Q$\times$K and Attention\_Prob $\times$V modules consume largest portions of energy and area since they are two most computational intensive modules. The latter consumes less energy thanks to local V pruning. \topk engines are relatively efficient, only taking \topkpowerratio\xspace of overall power and \topkarearatio\xspace of area so will not cause severe congestion issues.

\textbf{Comparisons with CPUs and GPUs.}
Figure~\ref{fig:perfcomp} shows the speedup and energy efficiency comparisons of \spattenfull\xspace with baselines on attention layers of the benchmarks. 
On average, \name achieves \perfovertitanxp\x, \perfoverxeon\x, \perfovernano\x, and \perfoverrasp\x speedup, and \eeovertitanxp\x, \eeoverxeon\x, \eeovernano\x, and \eeoverrasp\x energy saving over \titanxp, \xeon, \nano, and \rasp.
\name obtains high speedup because it has a highly parallelized and pipelined datapath. Meanwhile, cascade pruning and progressive \lowprec further reduce computation and DRAM access. Energy savings mainly come from DRAM fetch reduction. The specialized datapath also reduces intermediate SRAM fetch. 
Since FFN layer computations are reduced by token pruning, CPUs and GPUs can also be accelerated. 
We implement token pruning on CPUs/GPUs. We use \texttt{topk} and \texttt{gather} operations to select un-pruned tokens and QKV matrices to reduce matrix sizes, thus reducing computation, latency, and memory footprint. 3\x pruning ratio brings up to 2.3\x speedup for BERT in batch mode (assume performing token pruning twice). GPT-2 results in Figure~\ref{fig:perfcomp} do not have Beam Search. However, our techniques can also accelerate the Beam Search case because when a token (and its K, V) is pruned, it \emph{will not be used by any beams}.

\begin{table}[t]
\centering
\renewcommand*{\arraystretch}{0.1}
\setlength{\tabcolsep}{1pt}
\footnotesize
\centering
\caption{Compare \spattensmall with prior art \athree and \mnnfast.}
\vspace{-5pt}
\begin{tabular}{l|c|c|c}
\toprule
& \mnnfast & \athree & \textbf{\spattensmall}  \\
\midrule
\midrule
Cascade Head Pruning & \xmark & \xmark & \cmark \\
\midrule
Cascade Token Pruning & \xmark & \xmark & \cmark \\
\midrule
Interpretable Pruning & \xmark & \xmark & \cmark \\
\midrule
Local Value Pruning & \cmark & \cmark & \cmark \\
\midrule
Progressive Quantization & \xmark & \xmark & \cmark \\
\midrule
Preprocessing Overhead & \xmark & \cmark & \xmark \\
\midrule
Reduce Computation of & Attention only & Attention only & Attention and FFN \\
\midrule
Accelerate & \bert only & \bert only & \bert \& \gpttwo \\
\midrule
\midrule
Technology & FPGA (28nm) & ASIC (40nm) & ASIC (40nm)  \\
\midrule
Frequency & 1GHz (projected) & 1GHz & 1GHz \\
\midrule
Area (mm$^2$) & - & 2.08 mm$^2$ & 1.55 mm$^2$  \\
\midrule
Throughput (GOP/s) & 120 (1\x) & 221 (1.8\x) & 360 (\perfovermnnfast\x) \\
\midrule
Energy Effi. (GOP/j)
 & 120 (1$\times$) & 269 (2.2$\times$) & 382 (\eeovermnnfast$\times$) \\
\midrule
Area Effi. (GOP/s/mm$^2$)
& - & 106 (1$\times$) & 238 (2.2$\times$) \\
\bottomrule
\end{tabular}
\label{tab:compa3}
\vspace{-20pt}
\end{table}

\textbf{Comparisons with \athree and \mnnfast.}
\athree~\cite{ham20203} and \mnnfast~\cite{8980322} are also attention accelerators exploring sparsity. \athree first sorts each dimension of the key vectors among all keys. Then it uses a pre-specified number of largest/smallest elements in the keys to conduct multiplications with a query and get \emph{partial} attention scores. If a score is smaller than a threshold, then the corresponding key will be pruned. \mnnfast removes V vectors whose attention probabilities are smaller than a threshold.
We compare the differences and performance of \spattensmall, \athree, and \mnnfast in Table~\ref{tab:compa3}. Specifically: (i) In \athree and \mnnfast, all QKV vectors need to be fetched from DRAM to on-chip buffers before determining what can be pruned, so it cannot reduce DRAM access. Thus, they can only accelerate computation-bounded models (discriminative \bert), but cannot accelerate memory-bounded models (generative \gpttwo). (ii) \athree has pre-processing overhead -- sorting the keys. (iii) Token pruning in \name is \emph{global and cascade}, while that in \athree is local in one head. Therefore, only \name can reduce the computation in both attention and FFN layers. (iv) Cascade token pruning is interpretable and can be intuitively visualized step by step (Figure \ref{fig:pruning_examples}). (v) \name also supports head pruning and progressive quantization.
\begin{figure}[t]
    \centering
    \includegraphics[width=1\columnwidth]{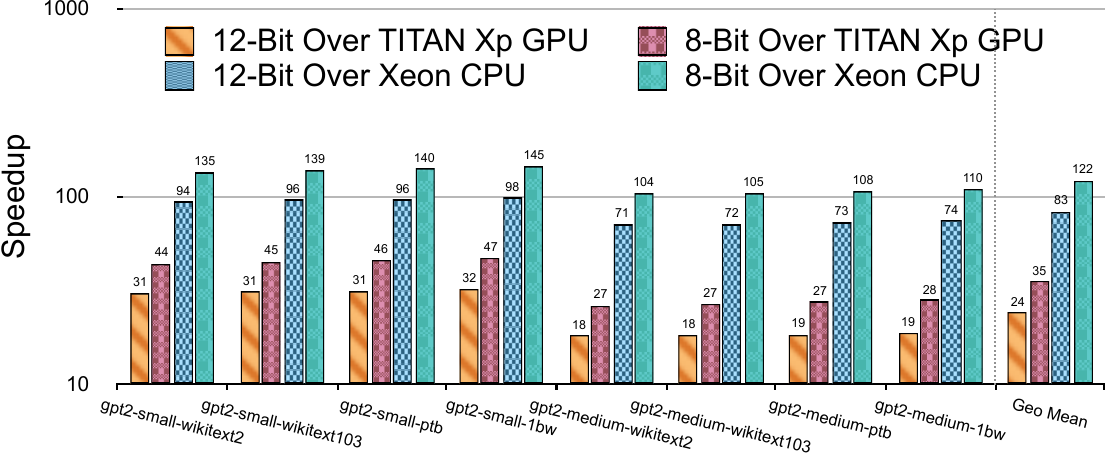}
    \vspace{-10pt}
    \caption{End-to-End speedup of {\spattenfc} over baselines. FC layer weights are quantized to 8 bits or 12 bits.}
    \vspace{-10pt}
    
    \label{fig:perfcompfc}
\end{figure}

\begin{table}[t]
\centering
\caption{FC \& Attention FLOPs \& Latency Breakdown on \gpttwo-Medium.}
\vspace{-5pt}
\setlength{\tabcolsep}{1pt}
\footnotesize
\centering
\begin{tabular}{l|c|c|c|c}
\toprule
& FC \gflops & Attn \gflops & FC Latency (ms) & Attn Latency (ms) \\
\midrule
GPU & 19.3 (85.6\%) & 3.3 (14.4\%) & 388.3 (51.4\%) & 366.7 (48.6\%) \\
\midrule
\textbf{\spattenfc} & 19.3 (95.5\%) & 0.9 (4.5\%) & 25.75 (92.4\%) & 2.13 (7.6\%)\\
\bottomrule

\end{tabular}
\label{tab:fc_attn}
\vspace{-10pt}
\end{table}

The parallelism $d$ in {\athree} is 64, corresponding to 128 multipliers. We compare \athree with {\spattensmall} which has the same number of multipliers (128), technology (40nm) and bandwidth (64GB/s) in Table~\ref{tab:compa3}.
Under 1GHz, \athree throughput is 2$\times d$=128\gflopss. Since it has 1.73\x geomean speedup, the effective throughput is 128$\times$1.72=221\gflopss. We include the same DRAM power for \athree and \spattensmall. \spattensmall\xspace achieves \perfoverathree\x better throughput, \eeoverathree\x better energy efficiency, and \aeoverathree\x better area efficiency over \athree.
\mnnfast essentially only supports local Value pruning. We get its throughput number with our reproduced simulator under the same bandwidth and number of multipliers. \mnnfast was originally a Zynq-7020 FPGA design (10W power). As an optimistic estimation, we assume ASIC implementation consumes 10\x less power, \ie 1W. Compared to \mnnfast, \spattensmall\xspace has \perfovermnnfast\x higher throughput, and \eeovermnnfast\x better energy efficiency.

\begin{figure}[t]
    \centering
    \includegraphics[width=\columnwidth]{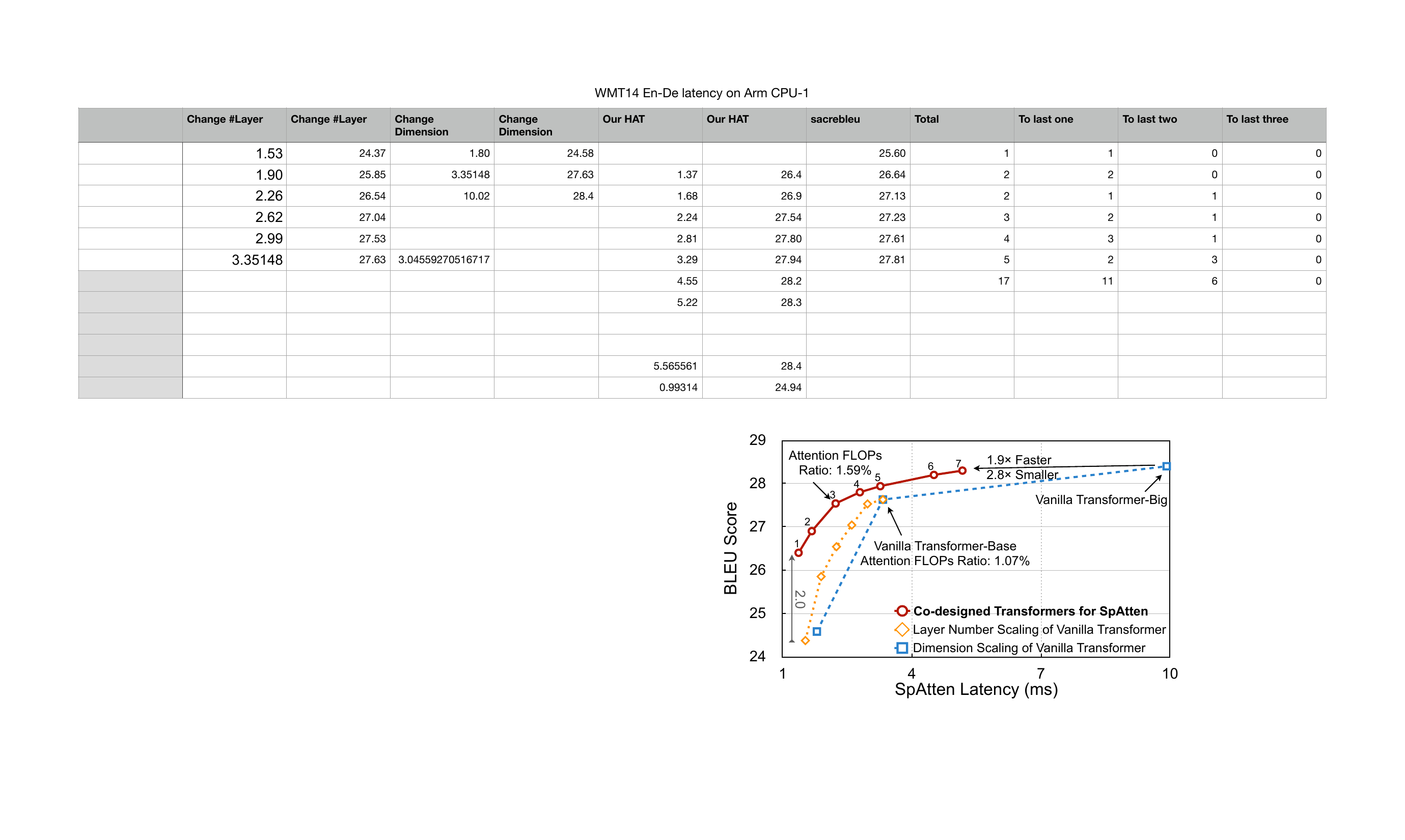}
    \vspace{-20pt}
    \caption{Co-designed Transformers for \spattenfc\xspace can achieve 1.9\x speedup and 2.8\x model size reduction over vanilla Transformers~\cite{46201}.
    }
    \label{fig:spattenhat}
    \vspace{-10pt}
\end{figure}

\textbf{End-to-End Performance with FFN Support.} To compare the end-to-end performance of \name with baselines. We extend our {\spattenfull} to support the FC in the  Feed-Forward Network (FFN) layers by reusing the multiplier arrays. The extended architecture is named {\spattenfc}.
FC weights are linear symmetrically quantized to 12 bits and 8 bits and stored on DRAM. Since the FCs in GPT-2 generation stage are matrix-vector multiplications, the performance of {\spattenfc} is memory-bounded. As shown in Figure~\ref{fig:perfcompfc}, on eight \gpttwo-Medium benchmarks, 8-bit FC \spattenfc\xspace achieves on-average \perffcovertitanxpteight\x and \perffcoverxeoneight\x speedup over \titanxp\xspace and \xeon; 12-bit FC \spattenfc\xspace achieves \perffcovertitanxptwelve\x and \perffcoverxeontwelve\x respectively. The breakdowns of computation and latency of FC and attention parts of four \gpttwo-Medium benchmarks averaged are shown in Table~\ref{tab:fc_attn}. Head pruning is not employed in this comparison. \spattenfc\xspace applies token pruning to reduce the attention FLOPs. The FC FLOPs are the same because token pruning can only reduce FC computation in the summarization stage (\bert), not the generation stage (\gpttwo). On GPU, the attention only accounts for 14.4\% computation but consumes 48.6\% latency, echoing with our analysis in Section~\ref{sec:motivation}. By contrast, attention on \spattenfc\xspace can be efficiently supported, thus only taking 7.6\% latency.

\begin{figure}[t]
    \centering
    \includegraphics[width=\columnwidth]{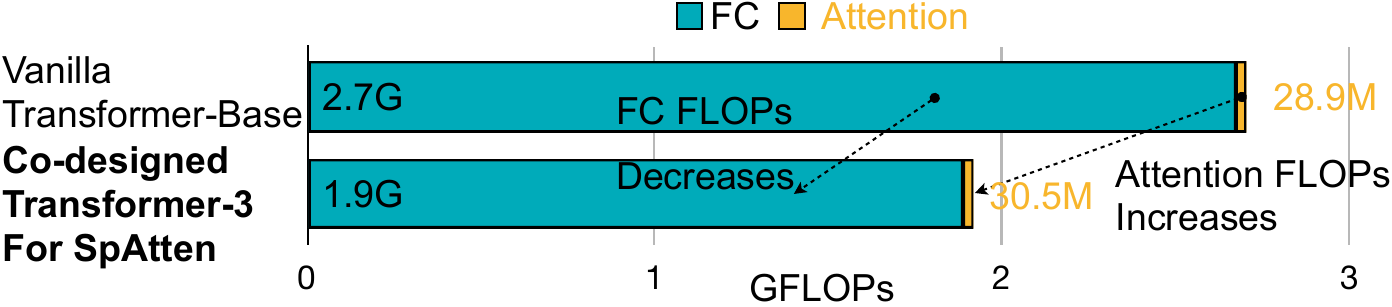}
    \vspace{-10pt}
    \caption{Co-designed Transformer has slightly more attention but much less FC computations because of \spattenfc's high efficiency on attention.
    }
    \label{fig:barhat}
    \vspace{-10pt}
\end{figure}

\textbf{Co-design Model Architecture with \name.} Besides the experiments above that leverage existing model architecture, we also explore the potentials of co-designing \name with model architecture by searching a 
Hardware-Aware Transformer (HAT)~\cite{hanruiwang2020hat} for \spattenfc. The search space contains [512, 640, 768] for embedding dim, [512, 1024, 2048, 3072] for FFN layer hidden dim, [1, 2, 3, 4, 5, 6] for decoder layer number, and last three layers for arbitrary encoder-decoder attention. Because the FC layers form the bottleneck of the \spattenfull\xspace performance, we intentionally configure the lower bound of FFN hidden dimension as low as 512 in expectation of reducing the FC ratio. We set different latency constraints and obtain a series of co-designed Transformers as shown in Figure~\ref{fig:spattenhat}. They are compared with layer number scaling and embedding dimension scaling of vanilla Transformer models~\cite{46201}. The co-designed Transformer-7 can achieve 1.9\x faster speed and 2.8\x smaller size over the vanilla Transformer-Big model. We also show the computation breakdowns of the vanilla Transformer-Base and the co-designed Transformer-3 in Figure~\ref{fig:barhat}. The two models have similar accuracy. Since \spattenfc\xspace can support attention with better efficiency, the co-designed model has a larger attention FLOPs. By virtue of the increased attention capacity, the FC computation can be largely shrunk without compromising the accuracy.

\subsection{Performance Analysis}
\begin{figure}[t]
    \centering
    \includegraphics[width=\columnwidth]{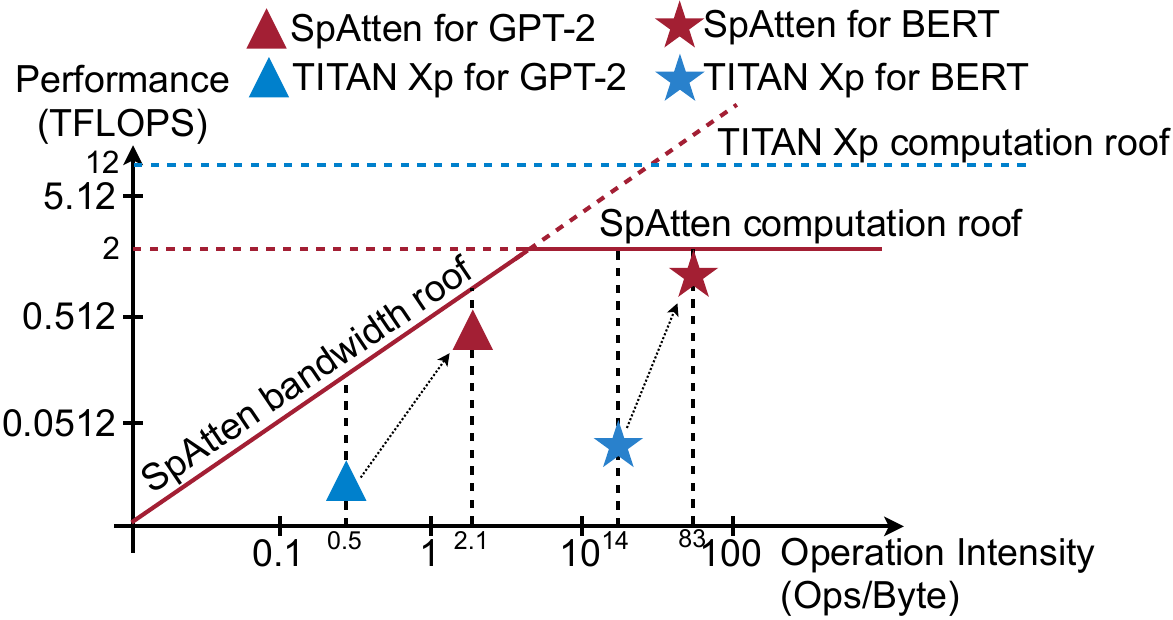}
    \vspace{-20pt}
    \caption{Roofline Model. The performance of \name is close to bandwidth and computation roofs.
    }
    \vspace{-15.5pt}
    \label{fig:roofline}
\end{figure}

\textbf{Roofline Analysis.}
To better understand the distance of \spattenfull\xspace to the theoretical optimal performance, we analyze its roofline model in Figure~\ref{fig:roofline} and compare it with \titanxp. We use theoretical operational intensity: only memory access for input QKV and attention outputs are counted. HBM has 512GB/s bandwidth; thus, the slope of the bandwidth roof is 512G. \spattenfull\xspace has 1024 multipliers; hence the theoretical computation roof (multiplication and addition) is 2\tflopss. For \bert, the operation intensity is high, so the performance is computation-bounded. \spattenfull\xspace achieves \spattenbertflops TFLOPS on \bert tasks. That is close to the computation roof and higher than GPU's \gpubertflops\tflopss. \gpttwo models, on the contrary, have a low arithmetic intensity and appear in the memory-bounded region. \spattenfull\xspace achieves \spattengpttwoflops\tflopss, close to the bandwidth roof and higher than GPU's \gpugpttwoflops\tflopss. GPU performance is far from the roofs in both models because of the low utilization of computation units. Progressive \lowprec improves the computation intensity; thus, the points of \spattenfull\xspace are to the right of GPU.

\begin{figure}[t]
    \centering
    \includegraphics[width=\columnwidth]{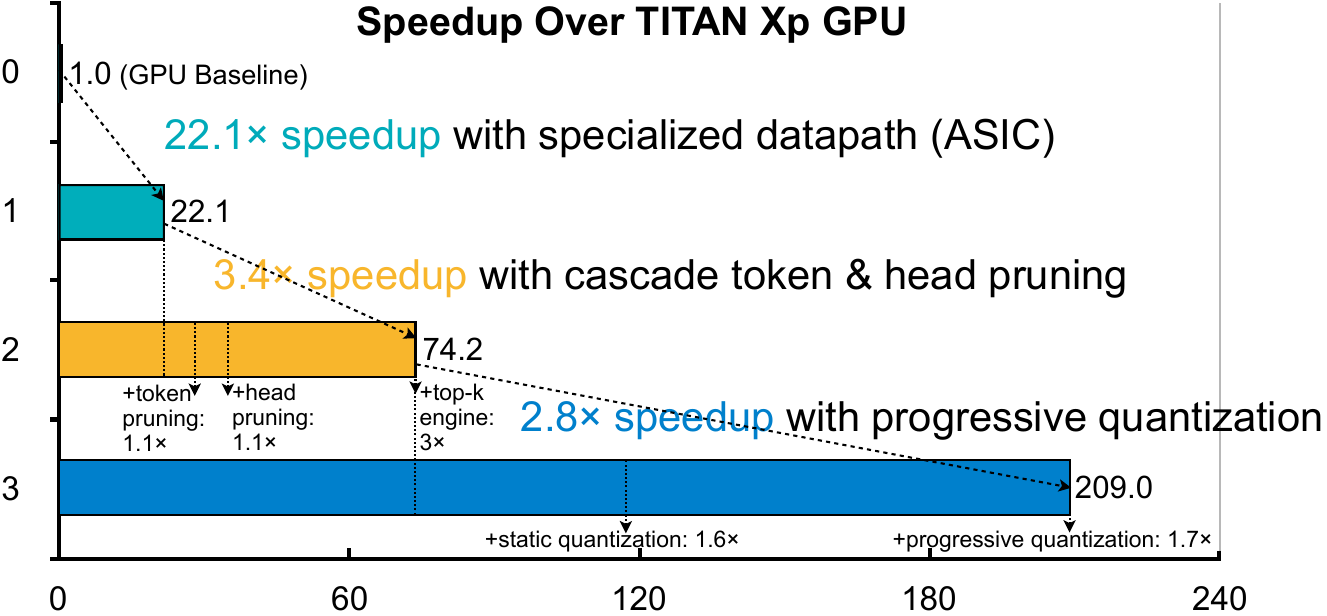}
     \vspace{-10pt}
    \caption{Speedup Breakdown of \name on \gpttwo models. Cascade pruning and progressive quantization bring 3.4\x and 2.8\x speedup, respectively.
    }
    \vspace{-5pt}
    
    \label{fig:perfbreak}
  
\end{figure}

\textbf{Breakdown of Speedup.}
Figure~\ref{fig:perfbreak} shows the speedup breakdown of \spattenfull\xspace over \titanxp \xspace on eight \gpttwo benchmarks. With a dedicated datapath, \spattenfull\xspace is 22.1$\times$ faster than GPU baseline, which needs to execute numerous memory instructions for attention. Cascade pruning is then applied to remove unimportant tokens and heads in the second step, reducing computation by \prunegpttwodramsaving\x and \hpdramsaving\x, respectively. However, the performance only improves by 1.1$\times$ for both. The reason is that cascade pruning needs to frequently execute \topk to find unimportant tokens/heads, which becomes a bottleneck without a high throughput \topk engine. Therefore, after adding the high-parallelism \topk engine, the bottleneck is resolved, and the performance jumps by 3\x. Finally, the progressive \lowprec reduces the average \bitwidth of inputs, achieving another 2.8$\times$ speedup with less DRAM access.

\begin{figure}[t]
    \centering
    \includegraphics[width=\columnwidth]{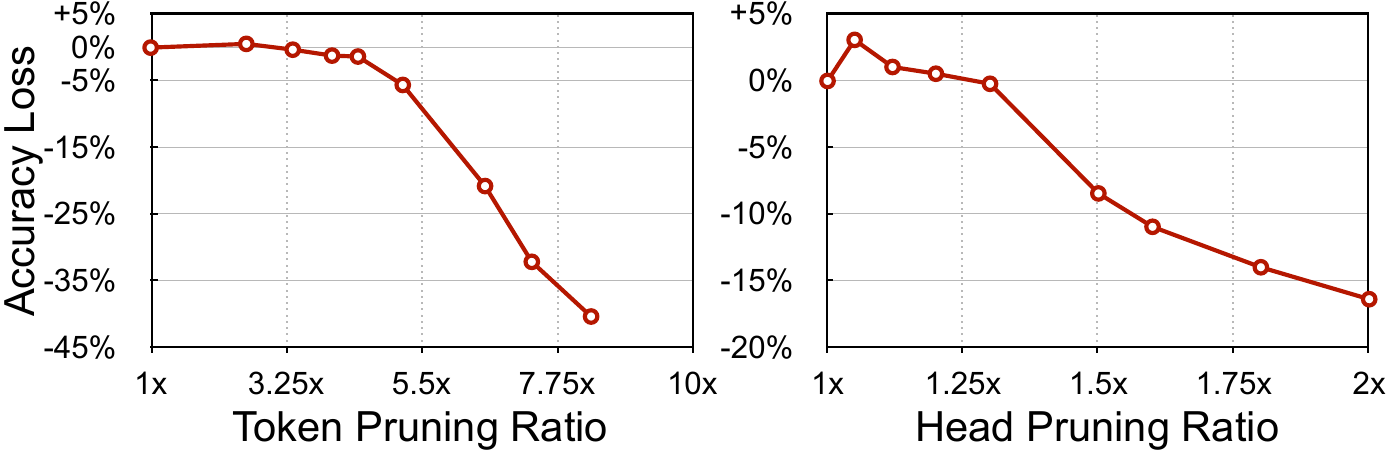}
    \vspace{-10pt}
    \caption{Trade-off curves between token/head pruning ratio and accuracy loss.}
    
    \label{fig:tradeoff}
    
    \vspace{-15pt}
\end{figure}

\textbf{Efficiency-Accuracy Trade-offs.}
Without accuracy loss, token pruning can prune \prunedramsaving\x for all benchmarks on average, while head pruning can prune \hpdramsaving\x. Figure~\ref{fig:tradeoff} shows two trade-off curves between the token/head pruning ratio and accuracy of GPT-2-Small on PTB (left) and BERT-Base on CoLA (right). For the token pruning curve, head pruning is not applied, and vice versa. We apply 12-bit quantization and disable progressive quantization for both. We can prune around 4$\times$ tokens for PTB and 1.2$\times$ heads for CoLA without accuracy loss. Small pruning ratios even improve the accuracy. Note that the pruning ratio is related to the input sentence length. Since the sentence length of \gpttwo benchmarks (around 1000) is much longer than \bert ones (less than 100), \gpttwo's pruning ratios can be larger while preserving the accuracy.

\begin{figure}[t]
    \centering
    \includegraphics[width=\columnwidth]{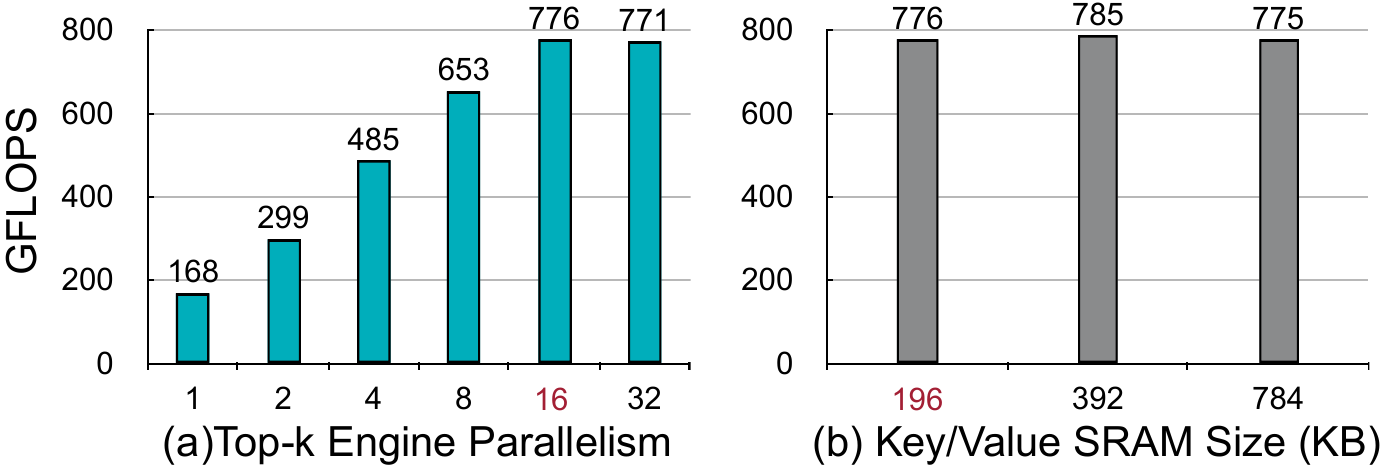}
    \vspace{-10pt}
    \caption{Design Space Exploration. A top-k engine with a parallelism of 16 and 196KB key/value buffer is sufficient for our settings.
    }
    \label{fig:ab2}
    \vspace{-10pt}
\end{figure}

\textbf{Design Choice Explorations.} We also explore the best architectural settings for \spattenfull\xspace in Figure~\ref{fig:ab2} on one \gpttwo application. The left side shows the performance with different parallelism (comparator number) of the \topk engine. 
Comparator number influences the time to perform a STATE\_RUN stage (see Algorithm~\ref{algorithm:topk}). 
After parallelism 16, the performance does not increase much because 16 matches the \topk engine input data rate from the Q$\times$K module. 
Thus parallelism larger than 16 makes \topk no longer the bottleneck and cannot much influence the overall performance. We select 16 in our design. 
On the right side, we change the size of SRAM storing K and V. Since \name supports up to 1024-length context, the minimum SRAM size is set to 2$\times$1024$\times$64$\times$12bits=196KB. `2' is for double buffering. Increasing SRAM size hardly increases the performance because the whole architecture is fully pipelined. More intermediate buffers will not significantly impact the throughput. 
Therefore, in consideration of reducing SRAM static power, we select the smallest 196KB.

\begin{figure*}[t]
    \centering
    \includegraphics[width=1\textwidth]{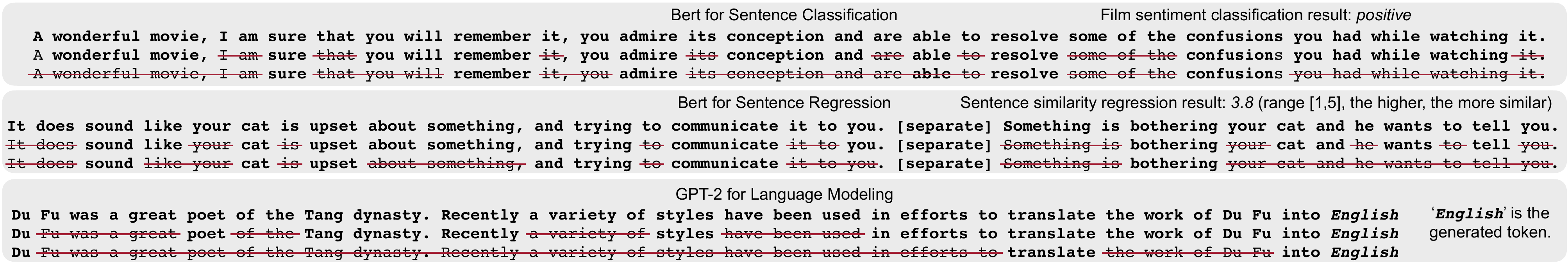}
    \vspace{-20pt}
    \caption{Examples for cascade token pruning in different models and tasks. \name supports both discriminative models (\bert) and generative models (\gpttwo). Unlike \athree and \mnnfast, cascade token pruning in \name is structured and interpretable.}
    \vspace{-20pt}
    \label{fig:pruning_examples}
\end{figure*}

\textbf{Interpretation and Visualization.}
Figure~\ref{fig:pruning_examples} visualizes the cascade token pruning process on various tasks. The pruned tokens are redundant ones such as `it, are, to, is', showing the effectiveness of \name's importance score mechanism. 
In the first example, the tokens that survive pruning are `remember', `admire', `resolve confusion'; we can easily interpret why the sentence is classified as a positive sentiment.
The second example is the similarity score regression. The regressed scores range from 1 to 5, and larger scores indicate higher similarity between two sentences. \name can effectively prune away the meaningless tokens such as `your' and `is', and keep the token pairs in two sentences such as `upset' and `bothering'.
The last example is a generative language modeling with \gpttwo. The generated token is `English'. \name aggressively prunes away most tokens as they are irrelevant to the generated token, and only keeps `Du', `translate' and `into' tokens. The model may find the name `Du' not typical in English, so the translation language should be `English'.

\begin{figure}[t]
    \centering
    \includegraphics[width=\columnwidth]{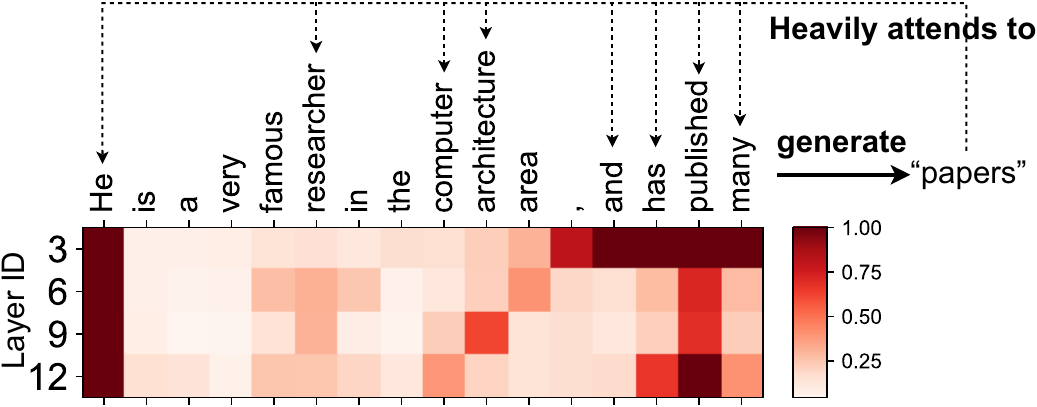}
    \vspace{-20pt}
    \caption{Cumulative importance scores in \gpttwo. Unimportant tokens are pruned on the fly. Important tokens are heavily attended.
}
\vspace{-15pt}
    \label{fig:heatmap_gpt2}
\end{figure}

Figure~\ref{fig:heatmap_gpt2} shows the cumulative importance scores of every single layer in a \gpttwo LM model. The important tokens are consistent across layers, such as `published'. Generated `papers' token heavily attends to several nearby tokens such as `published' and `many'. It also attends to some important tokens such as `researcher' and `architecture' even though they are far from it. In summary, token pruning reduces the model complexity and shows which tokens are attended most by the model, bringing better interpretability than \athree and \mnnfast.

\section{Related Work}
\subsection{Neural Networks Pruning and Quantization}
Neural networks tend to be over-parameterized, and many algorithms have been proposed to prune away the redundant parameters. Fine-grained pruning~\cite{han2016deep, wang2020apq} cut offs the connections within the weight matrix and achieves high pruning ratios. However, it is not friendly to CPUs and GPUs and requires dedicated hardware\cite{pal2018outerspace, sparch} to support sparse matrix multiplication, which may consume extra efforts for design and automation~\cite{wang2020gcn, mao2019park, wang2018learning}. To this end, structured pruning such as channel-pruning~\cite{he2017channel, he2018amc} was further proposed to remove the entire channel to enable acceleration on general-purpose hardware.~\cite{park2016faster} further proposed to enable fine-grained pruning speedup on general-purpose platforms. However, it requires a complicated guided sparsity learning process. Quantization reduces data precision to shrink the model size and bandwidth requirements. 
\name is fundamentally different from the existing weight pruning and quantization because there is no weight in attention, and we prune the input tokens/heads based on their importance to the sentence. Quantization in \name is applied to input QKV instead of weights.

\subsection{Accelerators for Sparse and Quantized Neural Networks}
There have been various domain-specific FPGA and ASIC accelerators proposed for neural networks~\cite{mubarik2020printed, ramanathan2020look, song2020vr, kim2020duplo, liu2020duet, mo2020tfe, weng2020dsagen, vilim2020gorgon, ke2020recnmp, evans2020jpeg, gupta2020deeprecsys, zhao2020smartexchange, baek2020multi, ji2019fpsa, ishida2020supernpu, gan2020ptolemy, he2020newton, ghodrati2020planaria, feng2020mesorasi, singh2020nebula, imani2020deep, shiflett2020pixel, gudaparthi2019wire, shao2019simba, akin2019zcomp, huang2019ecnn}. Many of them leverage the sparsity to improve performance~\cite{10.1109/ISCA.2016.30, zhang2016cambricon, parashar2017scnn, zhou2018cambricon, yang2019sparse, gondimalla2019sparten, sharify2019laconic, cong2018understanding, yang2020procrustes, gong2020save, hwang2020centaur, qin2020sigma, delmas2019bit, yan2020hygcn, song2020accpar, zhang2020enabling, hua2019boosting}.
There also exist general sparse tensor algebra accelerators~\cite{chen2020tpspmv, hegde2019extensor, kanellopoulos2019smash, zhu2019sparse, kwon2019tensordimm, mahmoud2020tensordash, srivastava2020matraptor, asgari2020alrescha, srivastava2020tensaurus, weng2020hybrid, lascorz2019shapeshifter, sadi2019efficient} proposed in recent years, which can be used to process sparse FC layers. 
Most of the prior work focuses on leveraging weight sparsity. By contrast, \name leverages activation (token/head) sparsity and employs specialized \topk engines to support on-the-fly cascade token/head pruning.
Quantized neural networks are also supported by many accelerators~\cite{lee2018unpu, judd2016stripes, sharify2018loom, umuroglu2018bismo, rzayev2017deeprecon, sharma2018bit, nvtensor, jouppi2017datacenter, song2020drq, zadeh2020gobo}. 
In those accelerators, the \bitwidth is fixed at the compile time, while in \name, we can adjust the \bitwidth according to the attention probability distributions.

\subsection{Efficient Natural Language Processing}
The large computation of attention-based NLP models raised much interest in improving their efficiency~\cite{hanruiwang2020hat, yan2020micronet, gu2018nonautoregressive, wang2020efficient}. GOBO~\cite{zadeh2020gobo} proposed to compress \bert model down to 3 bits, thus significantly reducing DRAM access. \powerbert~\cite{goyal2020power} prunes tokens based on the \emph{instant} attention probabilities of only one layer, which is different from \name's \emph{cumulative} attention probabilities of multiple layers. It cannot support \emph{per-head granularity} token pruning or local V vector pruning either. Head pruning is proposed in \cite{michel2019sixteen, voita2019analyzing} but they only prune head weights instead of activations as in \name. The head pruning in~\cite{voita2019analyzing} is not cascaded since pruned heads in one layer appear in latter layers. Our token pruning idea can also be generalized to Memory-Augmented Networks~\cite{NIPS2015_5846} to remove unimportant memory vectors and improve efficiency.

\section{Conclusion}

 We propose \name, a software-architecture \codesign to enable efficient sparse and quantized attention inference.
 We first propose cascade token and head pruning to remove the computation and memory access of inessential tokens and heads. 
 A novel \topk engine is designed to support on-the-fly token and head importance ranking with $O(n)$ time complexity. Moreover, we propose progressive \lowprec to allow different \bitwidths across layers. 
 \name achieves orders of magnitude speedup and energy savings over traditional platforms, and is \perfoverathree\x and \perfovermnnfast\x faster than \athree and \mnnfast. We also provide detailed performance analysis, breakdown of each technique, and design space explorations, offering insights to future NLP accelerator designs.

\section*{Acknowledgement}

Part of this work was supported under NSF CAREER Award \#1943349 and DARPA SDH program. We thank MIT Data Science and AI Lab (DSAIL) for supporting this research. We thank Joel Emer, Stephen Keckler, Mike O'Connor, Donghyuk Lee for inspiring discussions.


\bibliographystyle{IEEEtran}
\bibliography{main.bib}

\end{document}